\documentclass[english,10pt,twocolumn,twoside]{IEEEtran}
\usepackage[T1]{fontenc}
\usepackage[latin9]{inputenc}
\usepackage{float}
\usepackage{units}
\usepackage{bm}
\usepackage{amsmath}
\usepackage{amsthm}
\usepackage{amssymb}
\usepackage{graphicx}

\makeatletter

\providecommand{\tabularnewline}{\\}
\floatstyle{ruled}
\newfloat{algorithm}{tbp}{loa}
\providecommand{\algorithmname}{Algorithm}
\floatname{algorithm}{\protect\algorithmname}

\theoremstyle{plain}
\newtheorem{thm}{\protect\theoremname}
\theoremstyle{plain}
\newtheorem{lem}[thm]{\protect\lemmaname}

\usepackage{colortbl}
\usepackage{cite}

\@ifundefined{showcaptionsetup}{}{%
 \PassOptionsToPackage{caption=false}{subfig}}
\usepackage{subfig}
\makeatother

\usepackage{babel}
\providecommand{\lemmaname}{Lemma}
\providecommand{\theoremname}{Theorem}

\begin{document}

\title{Rician MIMO Channel- and Jamming-Aware Decision Fusion}

\author{D.~Ciuonzo,\IEEEmembership{~Senior~Member,~IEEE,} A.~Aubry,\IEEEmembership{~Senior~Member,~IEEE,}
and V.~Carotenuto,\IEEEmembership{~Member,~IEEE}\thanks{Manuscript received 20th November 2015; revised 3rd August 2016; accepted
23th February 2017. The associate editor coordinating the review of
this manuscript and approving it for publication was Prof. Xavier
Mestre.\protect \\
D. Ciuonzo was with University of Naples \textquotedbl{}Federico II\textquotedbl{},
DIETI, Via Claudio 21, 80125 Naples, Italy. He is now with Networking
Measurement and Monitoring (NM-2) s.r.l., 80143 Naples, Itally. (e-mail:
domenico.ciuonzo@ieee.org)\protect \\
A. Aubry and V. Carotenuto are with University of Naples \textquotedbl{}Federico
II\textquotedbl{}, DIETI, Via Claudio 21, 80125 Naples, Italy. (e-mail:
\{augusto.aubry, vincenzo.carotenuto\}@unina.it)}\vspace{-0.5cm}
}
\maketitle
\begin{abstract}
In this manuscript we study channel-aware decision fusion (DF) in
a wireless sensor network (WSN) where: ($i$) the sensors transmit
their decisions simultaneously for spectral efficiency purposes and
the DF center (DFC) is equipped with multiple antennas; ($ii$) each
sensor-DFC channel is described via a Rician model. As opposed to
the existing literature, in order to account for stringent energy
constraints in the WSN, only statistical channel information is assumed
for the non-line-of-sight (scattered) fading terms. For such a scenario,
sub-optimal fusion rules are developed in order to deal with the exponential
complexity of the likelihood ratio test (LRT) and impractical (complete)
system knowledge. Furthermore, the considered model is extended to
the case of (partially unknown) jamming-originated interference. Then
the obtained fusion rules are modified with the use of composite hypothesis
testing framework and generalized LRT. Coincidence and statistical
equivalence among them are also investigated under some relevant simplified
scenarios. Numerical results compare the proposed rules and highlight
their jamming-suppression capability.
\end{abstract}


\section{Introduction}

\subsection{Motivation and Related Literature}

\IEEEPARstart{D}{ecision} Fusion (DF) in a wireless sensor network
(WSN) consists in transmitting local decisions about an observed phenomenon
from sensors to a DF center (DFC) for a global decision, with the
intent of surveillance and/or anomaly detection \cite{Chen2006}.
Typically all the studies had been focused on a parallel access channels
(PACs) with instantaneous \cite{Chen2004,Lei2010} or statistical
channel-state information (CSI) \cite{Jiang2005a}, although some
recent works extended to the case of multiple access channels (MACs).
Adoption of a MAC in WSNs is clearly attractive because of its \emph{increased
spectral efficiency}. 

Distributed detection over MACs was first studied in \cite{Li2007},
where perfect compensation of the fading coefficients is assumed for
each sensor. Non-coherent modulation and censoring over PACs and MACs
were analyzed in \cite{Berger2009} with emphasis on processing gain
and combining loss. The same scenario was studied in \cite{Li2011},
focusing on the error exponents (obtained through the large deviation
principle) and the design of energy-efficient modulations for Rayleigh
and Rice fading. Optimality of received-energy statistic in Rayleigh
fading scenario was demonstrated for diversity MACs with non-identical
sensors in \cite{Ciuonzo2013}. Efficient DF over MACs only with knowledge
of the instantaneous channel gains and with the help of power-control
and phase-shifting techniques was studied in \cite{Umebayashi2012}.
Techniques borrowed from direct-sequence spread-spectrum systems were
combined with on-off keying (OOK) modulation and censoring for DF
in scenarios with statistical CSI \cite{Yiu2009}.

DF over a (virtual) MIMO (this setup will be referred to as ``MIMO-DF''
hereinafter) was first proposed in \cite{Zhang2008}, with focus on
power-allocation design based on instantaneous CSI, under the framework
of J-divergence. Distributed detection with ultra-wideband sensors
over MAC was then studied in \cite{Bai2010}. The same model was adopted
to study sensor fusion over MIMO channels with amplify-and-forward
sensors in \cite{Banavar2012,Nevat2014}. A recent theoretical study
on data fusion with amplify and forward sensors, Rayleigh fading channel
and a large-array at the DFC has been presented in \cite{Jiang2015}. 

Design of several sub-optimal fusion rules for MIMO-DF scenario was
given in \cite{Ciuonzo2012} in a setup with instantaneous CSI and
Rayleigh fading, while the analysis was extended in \cite{Ciuonzo2015}
to a large-array at the DFC, estimated CSI and inhomogeneous large
scale fading. In both cases binary-phase shift keying (BPSK) has been
employed. It is worth noticing that in MIMO-DF scenario the log-likelihood
ratio (LLR) is not a viable solution, since it suffers from the exponential
growth of the computational complexity with respect to (w.r.t.) the
number of sensors and a strong requirement on system knowledge.

However, frequently the final purpose of a WSN is anomaly detection
(viz. the null hypothesis is much more frequent than the alternative
hypothesis, denoting the ``anomaly''). Such problem arises in many
application contexts, such as intrusion detection or monitoring of
hazardous events. In this case, a wise choice of the modulation format
is the on-off keying (OOK), which ensures a nearly-optimal censoring
policy (and thus significant energy savings) \cite{Berger2009,Rago1996}.
Additionally, though the channel between the sensors and the DFC may
be accurately modelled as Rician, assuming instantaneous CSI (i.e.,
estimation of the scattered fading component) may be too energy costly
for an anomaly detection problem. This motivated the study of DF over
Rician MAC channels (in the single-antenna DFC case) with only statistical
CSI in \cite{Li2011,P.SalvoRossi2014}. We point out that statistical
CSI is instead a reasonable assumption for a WSN and can be obtained
through long-term training-based techniques (since statistical parameters
of Rician model have a slower variation with respect to the coherence
time of the channel), mimicking the procedures proposed in \cite{Dogandzic2005}.
The aforementioned problem may be further exacerbated by the presence
of a (possibly distributed) jamming device in the WSN deployment area
\cite{Xu2006}. Such problem is clearly relevant in non-friendly environments,
such as the battlefield, where malicious devices (i.e., the jammers)
are placed to hinder the operational requirements of the WSN. Indeed,
due to jammer hostile nature, unknown interference is superimposed
to useful received signal (containing ``informative'' sensors contributions).
Therefore, additional relevant parameters may be unknown at the DFC
side. This precludes development of sub-optimal (simplified) fusion
rules based on the LLR, which assume complete specification of pdfs
under both hypotheses. To the best of our knowledge, the study of
such a setup for MIMO-DF has not been addressed yet in the open literature.

\subsection{Main Results and Paper Organization}

The contributions of the present manuscript are summarized as follows.
\begin{itemize}
\item We study decision fusion over MAC with Rician fading and multiple
antennas at the DFC (as opposed to \cite{Li2011,P.SalvoRossi2014}).
In the present study only the LOS component is assumed known at the
DFC. Also, by adopting the same general assumptions in \cite{Ciuonzo2015},
the considered model also accounts for unequal long-term received
powers from the sensors, through a common path loss and shadowing
model;
\item We derive sub-optimal fusion rules dealing with exponential complexity
and with required system knowledge in the considered scenario, namely,
we derive ($i$) ``ideal sensors'' (IS) (following the same spirit
as in \cite{Chen2004,Ciuonzo2014a}), ($ii$) ``non-line of sight''
(NLOS), ($iii$) ``widely-linear'' (mimicking \cite{Ciuonzo2015})
and ($iv$) ``improper Gaussian moment matching'' (IGMM, based on
second order characterization of the received vector under both hypotheses)
rules; 
\item Subsequently, we consider DF in the presence of a (either distributed
or co-located) multi-antenna jamming device, whose communication channel
is described by an analogous Rician model. The problem is tackled
within a composite hypothesis testing framework and solved via the
generalized likelihood-ratio test (GLRT) \cite{Kay1998} and similar
key simplifying assumptions as in the ``no-jamming'' scenario, thus
leading to IS-GLRT, NLOS-GLRT and IGMM-GLRT rules, respectively;
\item Simulation studies (along with a detailed complexity analysis) are
performed to compare the performance of the considered rules and verify
the asymptotical equivalences (later proved in Secs. \ref{subsec: Asymptotic Eq no sub}
and \ref{subsec: Asymptotic Equivalences Jammer}) among them in some
specific instances. Also, the performance trend as a function of the
Rician parameters of the WSN and the jammer, the thermal noise and
the number of receive antennas are investigated and discussed.
\end{itemize}
The remainder of the manuscript is organized as follows: Sec.~\ref{sec:System-Model}
introduces the model; in Sec.~\ref{sec:Fusion-Rules} we derive and
study the fusion rules, while in Sec.~\ref{sec: Subspace Interference}
we generalize the analysis to the case of a subspace interference;
the obtained rules are compared in terms of computational complexity
in Sec. \ref{sec: Complexity analysis}; in Sec.~\ref{sec: Simulation results}
we compare the presented rules through simulations; finally in Sec.~\ref{sec: Conclusions}
we draw some conclusions; proofs and derivations are contained in
a dedicated Appendix.

\emph{Notation} - Lower-case (resp. Upper-case) bold letters denote
vectors (resp. matrices), with $a_{n}$ (resp. $a_{n,m}$) being the
$n$th (resp. the $(n,m)$th) element of $\bm{a}$ (resp. $\bm{A}$);
upper-case calligraphic letters denote finite sets, with $\mathcal{A}^{K}$
representing the $k$-ary Cartesian power of $\mathcal{A}$; $\bm{O}_{N\times K}$
(resp. $\bm{I}_{N}$) denotes the $N\times K$ (resp. $N\times N$)
null (resp. identity) matrix, with corresponding short-hand notation
$\bm{O}_{N}$ for a square matrix; $\bm{0}_{N}$ (resp. $\bm{1}_{N}$)
denotes the null (resp. ones) vector of length $N$; $\bm{a}_{n:m}$
(resp. $\bm{A}_{n:m}$) denotes the sub-vector of $\bm{a}$ (resp.
the sub-matrix of $\bm{A}$) obtained from selecting only $n$th to
$m$th elements of $\bm{v}$ (resp. $n$th to $m$th rows/columns
of $\bm{A}$); $\mathbb{E}\{\cdot\}$, $\mathrm{var\{\cdot\}}$, $(\cdot)^{T}$,
$(\cdot)^{\dagger}$, $(\cdot)^{-}$, $\Re\left(\cdot\right)$, $\Im(\cdot)$
and $\left\Vert \cdot\right\Vert $ denote expectation, variance,
transpose, conjugate transpose, pseudo-inverse, real part, imaginary
part and Euclidean norm operators, respectively; $(\cdot)_{+}$ is
used to indicate $\max\{0,\cdot\}$; $\mathrm{diag}(\bm{A})$ (resp.
$\mathrm{diag}(\bm{a})$) denotes the diagonal matrix extracted from
$\bm{A}$ (resp. the diagonal matrix with main diagonal given by $\bm{a}$);
$\mathrm{det(}\bm{A})$ is used to denote the determinant of $\bm{A}$;
$\mathrm{\lambda{}_{min}}(\bm{A})$ denotes the minimum eigenvalue
of the Hermitian matrix $\bm{A}$; $\bm{P}_{\bm{X}}^{\perp}$ denotes
the orthogonal projector to the range space spanned by $\bm{X}$;
$\underline{\bm{a}}$ (resp. $\underline{\bm{A}}$) denotes the augmented
vector (resp. matrix) of $\bm{a}$ (resp. $\bm{A}$), that is $\underline{\bm{a}}\triangleq\left[\begin{array}{cc}
\bm{a}^{T} & \bm{a}^{\dagger}\end{array}\right]^{T}$ (resp. $\underline{\bm{A}}\triangleq\left[\begin{array}{cc}
\bm{A}^{T} & \bm{A}^{\dagger}\end{array}\right]^{T}$); $P(\cdot)$ and $p(\cdot)$ denote probability mass functions (pmf)
and probability density functions (pdf), while $P(\cdot|\cdot)$ and
$p(\cdot|\cdot)$ their corresponding conditional counterparts; $\bm{\Sigma}_{\bm{x}}$
(resp. $\bar{\bm{\Sigma}}_{\bm{x}}$) denotes the covariance (resp.
the complementary covariance) matrix of the complex-valued random
vector $\bm{x}$; $\mathcal{N}_{\mathbb{C}}(\bm{\mu},\bm{\Sigma})$
(resp. $\mathcal{N}_{\mathbb{C}}(\bm{\mu},\bm{\Sigma},\bar{\bm{\Sigma}}$)
denotes a proper (resp. an improper) complex normal distribution with
mean vector $\bm{\mu}$ and covariance matrix $\bm{\Sigma}$ (resp.
covariance $\bm{\Sigma}$ and pseudo-covariance $\bar{\bm{\Sigma}}$),
while $\mathcal{N}(\bm{\mu},\bm{\Sigma})$ denotes the corresponding
real-valued counterpart; finally the symbols $\propto$, and $\sim$
mean ``statistically equivalent to'' and \textquotedblleft distributed
as\textquotedblright , respectively.

\section{System Model\label{sec:System-Model}}

Hereinafter we will consider a decentralized binary hypothesis test,
where $K$ sensors are used to discern between the hypotheses in the
set $\mathcal{H}\triangleq\{\mathcal{H}_{0},\mathcal{H}_{1}\}$ (e.g.
$\mathcal{H}_{0}/\mathcal{H}_{1}$ may represent the absence/presence
of a specific target of interest). The $k$th sensor, $k\in\mathcal{K}\triangleq\{1,2,\ldots,K\}$,
takes a binary local decision $\xi_{k}\in\mathcal{H}$ about the observed
phenomenon on the basis of its own measurements. Here we do not make
any conditional (given $\mathcal{H}_{i}\in\mathcal{H}$) mutual independence
assumption on $\xi_{k}$. Each decision $\xi_{k}$ is mapped to a
symbol $x_{k}\in{\cal X}=\{0,+1\}$ representing an OOK modulation:
without loss of generality (w.l.o.g.) we assume that $\xi_{k}=\mathcal{H}_{i}$
maps into $x_{k}=i$, $i\in\{0,1\}$. The quality of the WSN is characterized
by the conditional joint pmfs $P(\bm{x}|\mathcal{H}_{i})$. Also,
we denote $P_{D,k}\triangleq P\left(x_{k}=1|\mathcal{H}_{1}\right)$
and $P_{F,k}\triangleq P\left(x_{k}=1|\mathcal{H}_{0}\right)$ the
probability of detection and false alarm of the $k$th sensor, respectively
(here we make the assumption $P_{D,k}\geq P_{F,k}$, meaning that
each sensor decision procedure leads to receive operating characteristics
above the chance line \cite{Scharf1991}). In some situations, aiming
at improving clarity of exposition, we will use the short-hand notation
$(P_{0,k},P_{1,k})=(P_{F,k},P_{D,k})$ (and $(P_{0},P_{1})=(P_{F},P_{D})$,
in the simpler case of conditionally i.i.d. decisions).

Sensors communicate with a DFC equipped with $N$ receive antennas
over a wireless flat-fading MAC in order to exploit diversity so as
to mitigate small-scale fading; this setup determines a \emph{distributed
(or virtual) MIMO}\textbf{ }channel \cite{Zhang2008,Ciuonzo2012}.
Also, perfect synchronization\footnote{Multiple antennas at the DFC do not make these assumptions harder
to verify w.r.t. a single-antenna MAC.}, as in \cite{Li2007,Ciuonzo2013,Zhang2008,Ciuonzo2012}, is assumed
at the DFC.

We denote: $y_{n}$ the signal at the $n$th receive antenna of the
DFC after matched filtering and sampling; $(\sqrt{d_{k,k}}\,\bar{h}_{n,k})$
the composite channel coefficient between the $k$th sensor and the
$n$th receive antenna of the DFC; $w_{n}$ the additive white Gaussian
noise at the $n$th receive antenna of the DFC. The vector model at
the DFC is:
\begin{align}
\bm{y}=\,\, & \bar{\bm{H}}\bm{D}^{1/2}\bm{x}+\bm{w}\label{eq:Channel_model}
\end{align}
where $\bm{y}\in\mathbb{C}^{N}$, $\bm{x}\in\mathcal{X}^{K}$ and
$\bm{w}\sim\mathcal{N}_{\mathbb{C}}(\bm{0}_{N},\sigma_{w}^{2}\bm{I}_{N})$
are the received-signal vector, the transmitted-signal vector and
the noise vector, respectively. Also, the matrices $\bar{\bm{H}}\in\mathbb{C}^{N\times K}$
and $\bm{D}\in\mathbb{C}^{K\times K}$ model independent small-scale
fading, geometric attenuation and log-normal shadowing. More specifically,
$\bm{D}\triangleq\mathrm{diag}\left(\begin{bmatrix}\beta_{1} & \cdots & \beta_{K}\end{bmatrix}^{T}\right)$
is a (known) matrix with $k$th diagonal element $d_{k,k}=\beta_{k}$
($\beta_{k}>0$) accounting for path loss and shadow fading experienced
by $k$th sensor. On the other hand, the $k$th column of $\bar{\bm{H}}$
models the (small-scale) fading vector of $k$th sensor as $\bar{\bm{h}}_{k}=b_{k}\,\bm{a}(\theta_{k})+\sqrt{1-b_{k}^{2}}\,\bm{h}_{k}$.
Here $\bm{a}(\cdot)$ denotes the steering vector (which depends on
the angle-of-arrival $\theta_{k}$, assumed known at the DFC\footnote{W.l.o.g. in this work we adopt a 1-D functional dependence for $\bm{a}(\cdot)$
(obtained from a ``far-field'' assumption), albeit more complicate
expressions could be considered as well.}) corresponding to the LOS component and $\bm{h}_{k}\sim\mathcal{N}_{\mathbb{C}}(\bm{0}_{N},\bm{I}_{N})$
corresponds to the normalized NLOS (scattered) component. Finally,
we denote $b_{k}\triangleq\sqrt{\frac{\kappa_{k}}{1+\kappa_{k}}}$,
where $\kappa_{k}$ represents the (known) usual Rician factor between
$k$th sensor and DFC.

The matrix $\bar{\bm{H}}$ can be expressed compactly in terms of
the relevant matrices $\bm{A}(\bm{\theta})\triangleq\begin{bmatrix}\bm{a}(\theta_{1}) & \cdots & \bm{a}(\theta_{K})\end{bmatrix}$,
$\bm{H}\triangleq\begin{bmatrix}\bm{h}_{1} & \cdots & \bm{h}_{K}\end{bmatrix}$
and $\bm{R}\triangleq\mathrm{diag}\left(\begin{bmatrix}b_{1} & \cdots & b_{K}\end{bmatrix}^{T}\right)$,
respectively, as
\begin{equation}
\bar{\bm{H}}\triangleq\,\,\bm{A}(\bm{\theta})\,\bm{R}+\bm{H}\,(\bm{I}_{K}-\bm{R}^{2})^{1/2}\,.\label{eq: H_bar}
\end{equation}
Finally, we underline that the \emph{received} scattered term from
$k$th sensor in Eq. (\ref{eq:Channel_model}) is $\sqrt{(1-b_{k}^{2})\,\beta_{k}}\,\bm{h}_{k}\sim\mathcal{N}_{\mathbb{C}}(\bm{0}_{N},\nu_{k}\,\bm{I}_{N})$,
where $\nu_{k}\triangleq[\beta_{k}\,(1-b_{k}^{2})]$, while its LOS
term is $\bm{\mu}_{k}\triangleq[\sqrt{\beta_{k}}\,b_{k}\,\bm{a}(\theta_{k})]$
and corresponds to the $k$th column of the matrix 
\begin{equation}
\widetilde{\bm{A}}(\bm{\theta})\triangleq\left(\bm{A}(\bm{\theta})\,\bm{R}\,\bm{D}^{1/2}\right)\,,
\end{equation}
denoting the matrix of \emph{received} LOS terms from the WSN.

\section{Fusion Rules\label{sec:Fusion-Rules}}

\subsection{Optimum (LLR) Rule }

The optimal test \cite{Kay1998} is formulated on the basis of the
LLR $\Lambda_{\mathrm{opt}}\triangleq\ln\left[\frac{p(\bm{y}|\mathcal{H}_{1})}{p(\bm{y}|\mathcal{H}_{0})}\right]$,
and decides in favour of $\mathcal{H}_{1}$ (resp. $\mathcal{H}_{0}$)
when $\Lambda_{\mathrm{opt}}>\gamma$ (resp. $\Lambda_{\mathrm{opt}}\leq\gamma$),
with $\gamma$ denoting the threshold which the LLR is compared to\footnote{ The threshold $\gamma$ can be determined to ensure a fixed system
false-alarm rate (Neyman-Pearson approach), or can be chosen to minimize
the probability of error (Bayesian approach) \cite{Kay1998}.}. After few manipulations, the LLR can be expressed explicitly as
\begin{gather}
\Lambda_{\mathrm{opt}}=\ln\left[\frac{\sum_{\bm{x}\in\mathcal{X}^{K}}\frac{P(\bm{x}|\mathcal{H}_{1})}{[\sigma_{e}^{2}(\bm{x})]^{N}}\exp(-\frac{\bm{\|y}-\sum_{k=1}^{K}\bm{\mu}_{k}\,x_{k}\|^{2}}{\sigma_{e}^{2}(\bm{x})})}{\sum_{\bm{x}\in\mathcal{X}^{K}}\frac{P(\bm{x}|\mathcal{H}_{0})}{[\sigma_{e}^{2}(\bm{x})]^{N}}\exp(-\frac{\bm{\|y}-\sum_{k=1}^{K}\bm{\mu}_{k}\,x_{k}\|^{2}}{\sigma_{e}^{2}(\bm{x})})}\right]\label{eq:optimum_llr}
\end{gather}
where $\sigma_{e}^{2}(\bm{x})\triangleq(\sigma_{w}^{2}+\sum_{k=1}^{K}\nu_{k}\,x_{k})$.
The above result follows from $\bm{y}|\mathcal{H}_{i}$ being a Gaussian
mixture random vector, since the pdf under each hypothesis can be
obtained as $p(\bm{y}|\mathcal{H}_{i})=\sum_{\bm{x}}p(\bm{y}|\bm{x})\,P(\bm{x}|\mathcal{H}_{i})$
(the directed triple $\mathcal{H}\rightarrow\bm{x}\rightarrow\bm{y}$
satisfies the Markov property). It is apparent that implementation
of Eq. (\ref{eq:optimum_llr}) requires a computational complexity
which grows exponentially with $K$ (namely $\mathcal{O}(2^{K})$,
where $\mathcal{O}(\cdot)$ stands for the usual Landau's notation).
Also, differently from \cite{Ciuonzo2012,Ciuonzo2015} (where a BPSK
modulation is employed), the computation here is complicated by the
fact that each component of the mixture (under $\mathcal{H}_{i}$)
has both different mean vectors and covariance (actually scaled identity)
matrices. Therefore, sub-optimal fusion rules with reduced complexity
are investigated in what follows.

\subsection{Ideal sensors (IS) rule\label{subsec: IS}}

The LLR in Eq.~(\ref{eq:optimum_llr}) can be simplified under the
assumption of perfect sensors \cite{Lei2010,Ciuonzo2012,Ciuonzo2013a},
i.e., $P(\bm{x}=\bm{1}_{K}|\mathcal{H}_{1})=P(\bm{x}=\bm{0}_{K}|\mathcal{H}_{0})=1$.
In this case $\bm{x}\in\{\bm{0}_{K},\bm{1}_{K}\}$ and Eq.~(\ref{eq:optimum_llr})
reduces to \cite{Ciuonzo2012}:
\begin{gather}
\ln\left[\frac{[\sigma_{e}^{2}(\bm{1}_{K})]^{-N}\,\exp\left(-\frac{\bm{\|y}-\sum_{k=1}^{K}\bm{\mu}_{k}\|^{2}}{\sigma_{e}^{2}(\bm{1}_{K})}\right)}{[\sigma_{w}^{2}]^{-N}\,\exp\left(-\frac{\bm{\|y}\|^{2}}{\sigma_{w}^{2}}\right)}\right]\propto\nonumber \\
2\,\Re(\bar{\bm{\mu}}^{\dagger}\bm{y})+\frac{\bar{\nu}}{\sigma_{w}^{2}}\left\Vert \bm{y}\right\Vert ^{2}\triangleq\Lambda_{{\scriptscriptstyle \mathrm{IS}}}\label{eq: IS rule}
\end{gather}
where $\bar{\bm{\mu}}\triangleq\frac{1}{K}\sum_{k=1}^{K}\bm{\mu}_{k}$
and $\bar{\nu}\triangleq\frac{1}{K}\sum_{k=1}^{K}\nu_{k}$ and terms
independent from $\bm{y}$ have been discarded (as they can be incorporated
in a suitably modified threshold $\gamma$).

It is worth noticing that the assumption of perfect local decisions
is used \emph{only} for system design purposes, and does \emph{not}
mean that the system is working under such ideal conditions, thus
the rule is suboptimal. Also, we observe that IS rule in (\ref{eq: IS rule})
is formed by a weighted combination of a maximum ratio combiner (MRC,
which actually is the statistic resulting from IS assumption on the
known part of channel vector at the DFC \cite{Ciuonzo2012}) and an
energy detector (ED, i.e., the statistic arising from the IS assumption
on the random part of the channel vector at the DFC \cite{Ciuonzo2013}).
Clearly, from Eq. (\ref{eq: IS rule}) it is apparent that IS rule
does not require sensor performance (i.e., the pmf $P(\bm{x}|\mathcal{H}_{i})$,
$i\in0,1$) for its implementation.

\subsection{Non line-of-sight (NLOS) rule\label{subsec: NLOS rule}}

In this case we derive a sub-optimal rule arising from the simplifying
assumption $\kappa_{k}=0$ (i.e., no sensor has a LOS path), thus
leading to: 
\begin{equation}
\ln\left[\frac{\sum_{\bm{x}\in\mathcal{X}^{K}}\frac{P(\bm{x}|\mathcal{H}_{1})}{[\bar{\sigma}_{e}^{2}(\bm{x})]^{N}}\exp(-\frac{\bm{\|y}\|^{2}}{\bar{\sigma}_{e}^{2}(\bm{x})})}{\sum_{\bm{x}\in\mathcal{X}^{K}}\frac{P(\bm{x}|\mathcal{H}_{0})}{[\bar{\sigma}_{e}^{2}(\bm{x})]^{N}}\exp(-\frac{\bm{\|y}\|^{2}}{\bar{\sigma}_{e}^{2}(\bm{x})})}\right]\label{eq: Energy test}
\end{equation}
where we have denoted $\bar{\sigma}_{e}^{2}(\bm{x})\triangleq(\sigma_{w}^{2}+\sum_{k=1}^{K}\beta_{k}x_{k})$.
We observe that in this case the LLR is function of the sole sufficient
statistic $\Lambda_{\mathrm{{\scriptscriptstyle NL}}}\triangleq\bm{\|y}\|^{2}$,
i.e., the energy of the received signal, which we retain as a simple
statistic for our test\footnote{We recall that, as in the case of IS rule, NLOS assumption is only
exploited at the design stage for development of simplified rule $\Lambda_{\mathrm{{\scriptscriptstyle NL}}}$.}. There is a twofold motivation for this choice. First, it was shown
in \cite{Ciuonzo2013} that under identical $\beta_{k}$'s and conditionally
independent decisions, the LLR in Eq. (\ref{eq: Energy test}) is
a monotone function of $\bm{\|y}\|^{2}$ (thus $\bm{\|y}\|^{2}>\gamma$
is the uniformly most powerful test \cite{Scharf1991}). Secondly,
by applying Gaussian moment matching to the simplified model in Eq.
(\ref{eq: Energy test}), the same test would be obtained. Therefore,
though we have no optimality claims for $\Lambda_{\mathrm{{\scriptscriptstyle NL}}}$
in this general case, we will consider NLOS rule as the decision statistic
due to its simplicity (and no requirements on sensors performance).

\subsection{Widely-linear (WL) rules}

It can be shown that $\bm{y}|\mathcal{H}_{i}$ has the following statistical
characterization up to the first two order moments (the proof is given
in Appendix):
\begin{align}
\mathbb{E}\{\bm{y}|\mathcal{H}_{i}\} & =\widetilde{\bm{A}}(\bm{\theta})\,\bm{\rho}_{i}\label{eq: 2nd order char MEAN}\\
\bm{\Sigma}_{\bm{y}|\mathcal{H}_{i}} & =\widetilde{\bm{A}}(\bm{\theta})\,\bm{\Sigma}_{\bm{x}|\mathcal{H}_{i}}\,\widetilde{\bm{A}}(\bm{\theta})^{\dagger}+\sigma_{e,i}^{2}\,\bm{I}_{N}\label{eq: 2nd order char COV}\\
\bar{\bm{\Sigma}}_{\bm{y}|\mathcal{H}_{i}} & =\widetilde{\bm{A}}(\bm{\theta})\,\bm{\Sigma}_{\bm{x}|\mathcal{H}_{i}}\,\widetilde{\bm{A}}(\bm{\theta})^{T}\label{eq: 2nd order char PCOV}
\end{align}
where $\bm{\rho}_{i}\triangleq\begin{bmatrix}P_{i,1} & \cdots & P_{i,K}\end{bmatrix}^{T}$
and $\sigma_{e,i}^{2}\triangleq[\sum_{k=1}^{K}\nu_{k}\,P_{i,k}+\sigma_{w}^{2}]$.
Therefore a convenient and effective approach consists in adopting
a WL statistic \cite{Schreier2010}. The WL approach (i.e., based
on the augmented vector $\underline{\bm{y}}$) is motivated by linear
complexity and $\bm{y}|\mathcal{H}_{i}$ being an \emph{improper }(cf.
Eq. (\ref{eq: 2nd order char PCOV})) complex-valued random vector,
that is $\bar{\bm{\Sigma}}_{\bm{y}|\mathcal{H}_{i}}\neq\bm{O}_{N}$.
More specifically, WL statistic is generically expressed as:
\begin{equation}
\Lambda_{{\scriptscriptstyle \mathrm{WL}}}\triangleq\underline{\bm{z}}^{\dagger}\bm{\underline{y}}\,,\label{eq: WL generical}
\end{equation}
 where the augmented vector $\underline{\bm{z}}$ has to be designed
according to a reasonable criterion. Then, $\Lambda_{{\scriptscriptstyle \mathrm{WL}}}$
is compared to a proper threshold $\gamma$ to obtain the corresponding
test.

Clearly, several optimization metrics may be considered for obtaining
$\underline{\bm{z}}$. The best choice (in a Neyman-Pearson sense)
would be searching for the WL rule maximizing the global detection
probability subject to a global false-alarm rate constraint, as proposed
in \cite{Quan2010} for a distributed detection problem. Unfortunately,
the optimized $\underline{\bm{z}}$ presents the following \emph{drawbacks}:
($i$) it is \emph{not} in \emph{closed-form}, ($ii$) it requires
a non-trivial optimization and ($iii$) it depends on the prescribed
false-alarm constraint. Additionally, the problem under investigation
is not a multivariate Gauss-Gauss test (i.e., $\bm{y}|\mathcal{H}_{i}\sim\mathcal{N}_{\mathbb{C}}(\bm{\mu}_{i},\bm{\Sigma}_{i})$)
but one discerning between mixtures of complex GMs (cf. Eq. (\ref{eq:optimum_llr})).
This would further complicate the optimization problem tackled in
\cite{Quan2010}.

Differently, in this paper we choose $\underline{\bm{z}}$ as the
maximizer of either the normal \cite{Picinbono1995} or modified \cite{Quan2008}
\emph{deflection measures}, denoted as $D_{0}(\,\underline{\bm{z}}\,)$
and $D_{1}(\,\underline{\bm{z}}\,)$ respectively, that is:
\begin{gather}
\underline{\bm{z}}_{\,{\scriptscriptstyle \mathrm{WL}},i}\triangleq\arg\max_{\underline{\bm{z}}:\,\left\Vert \,\underline{\bm{z}}\,\right\Vert ^{2}=1}D_{i}\left(\,\underline{\bm{z}}\,\right)\label{eq: Deflection max problem}\\
\mathrm{where}\quad D_{i}\left(\,\underline{\bm{z}}\,\right)\triangleq\frac{\left(\mathbb{E}\{\Lambda_{{\scriptscriptstyle \mathrm{WL}}}|\mathcal{H}_{1}\}-\mathbb{E}\{\Lambda_{{\scriptscriptstyle \mathrm{WL}}}|\mathcal{H}_{0}\}\right)^{2}}{\mathrm{var}\{\Lambda_{{\scriptscriptstyle \mathrm{WL}}}|\mathcal{H}_{i}\}}\nonumber 
\end{gather}
Maximization of deflection measures is commonly used in the design
of (widely) linear rules for DF, since $\underline{\bm{z}}_{\,{\scriptscriptstyle \mathrm{WL}},i}$
always admits a closed-form and also literature has shown acceptable
performance loss w.r.t. the LLR in analogous DF setups \cite{Quan2010,Lai2010}.
The vector $\underline{\bm{z}}_{\,{\scriptscriptstyle \mathrm{WL}},i}$,
being the optimal solution to the optimization in Eq. (\ref{eq: Deflection max problem}),
is (a similar proof can be found in \cite{Ciuonzo2015}):
\begin{gather}
\underline{\bm{z}}_{\,{\scriptscriptstyle \mathrm{WL}},i}=\frac{\bm{\Sigma}_{\underline{\bm{y}}|\mathcal{H}_{i}}^{-1}\,\underline{\widetilde{\bm{A}}}(\bm{\theta})\,\bm{\rho}_{1,0}}{||\bm{\Sigma}_{\underline{\bm{y}}|\mathcal{H}_{i}}^{-1}\,\underline{\widetilde{\bm{A}}}(\bm{\theta})\,\bm{\rho}_{1,0}||}\label{eq: Deflection maximizer WL formula}
\end{gather}
where $\bm{\rho}_{1,0}\triangleq(\bm{\rho}_{1}-\bm{\rho}_{0})$ and
$\bm{\Sigma}_{\underline{\bm{y}}|\mathcal{H}_{i}}$ is given by:
\begin{gather}
\bm{\Sigma}_{\underline{\bm{y}}|\mathcal{H}_{i}}=\underline{\widetilde{\bm{A}}}(\bm{\theta})\,\bm{\Sigma}_{\bm{x}|\mathcal{H}_{i}}\,\underline{\widetilde{\bm{A}}}(\bm{\theta})^{\dagger}+\sigma_{e,i}^{2}\,\bm{I}_{2N}\label{eq: Augmented covariance}
\end{gather}
The WL statistics are thus obtained employing Eq. (\ref{eq: Deflection maximizer WL formula})
into (\ref{eq: WL generical}). It is worth pointing out that, from
inspection of Eq. (\ref{eq: Deflection maximizer WL formula}), WL
rules only require knowledge up to the second order of the vectors
$\bm{x}|\mathcal{H}_{i}$.

\subsection{Improper Gaussian moment matching (IGMM) rule}

Differently here we fully exploit the second order characterization
provided in Eqs. (\ref{eq: 2nd order char MEAN}-\ref{eq: 2nd order char PCOV}).
In fact, after fitting $\bm{y}|\mathcal{H}_{i}$ to an improper complex
Gaussian, the following quadratic test can be obtained \cite{Schreier2010}:
\begin{eqnarray}
\Lambda_{\mathrm{{\scriptscriptstyle IGMM}}} & \triangleq & -(\underline{\bm{y}}-\mathbb{E}\{\underline{\bm{y}}|\mathcal{H}_{1}\})^{\dagger}\,\bm{\Sigma}_{\underline{\bm{y}}|\mathcal{H}_{1}}^{-1}\,(\underline{\bm{y}}-\mathbb{E}\{\underline{\bm{y}}|\mathcal{H}_{1}\})+\nonumber \\
 &  & (\underline{\bm{y}}-\mathbb{E}\{\underline{\bm{y}}|\mathcal{H}_{0}\})^{\dagger}\,\bm{\Sigma}_{\underline{\bm{y}}|\mathcal{H}_{0}}^{-1}\,(\underline{\bm{y}}-\mathbb{E}\{\underline{\bm{y}}|\mathcal{H}_{0}\})\label{eq: IGMM}
\end{eqnarray}
where $\mathbb{E}\{\underline{\bm{y}}|\mathcal{H}_{i}\}=\underline{\widetilde{\bm{A}}}(\bm{\theta})\,\bm{\rho}_{i}$
and $\bm{\Sigma}_{\underline{\bm{y}}|\mathcal{H}_{i}}$ is given in
Eq.~(\ref{eq: Augmented covariance}). IGMM rule presents the same
(reduced) requirements on knowledge of sensors performance as the
WL rules (cf. Eqs.~(\ref{eq: Deflection maximizer WL formula}) and
(\ref{eq: IGMM})). Differently, we expect it to perform nearly-optimal
(i.e., close to the LLR) at low SNR, as in such case both Gaussian
mixtures are well-approximated by a single Gaussian pdf.

\subsection{Asymptotic equivalences\label{subsec: Asymptotic Eq no sub}}

In this sub-section, we will establish asymptotic equivalences among
the proposed rules in the form of the following lemmas. These will
be employed as useful tools to facilitate the understanding of the
numerical comparisons shown in Sec.~\ref{sec: Simulation results}.
\begin{lem}
\label{lem: Exact NLOS}As the sensors approach a NLOS condition (i.e.,
the Rician factor $\kappa_{k}\rightarrow0$) IS and IGMM (recalling
$P_{D,k}\geq P_{F,k}$) rules are statistically equivalent to the
NLOS rule, i.e., they collapse to an energy detection test.
\end{lem}
\begin{IEEEproof}
The proof is obtained by substituting $\kappa_{k}=0$ in Eqs. (\ref{eq: IS rule})
and (\ref{eq: IGMM}), which respectively gives $\Lambda_{{\scriptscriptstyle \mathrm{IS}}}=\bm{\|y}\|^{2}(\sum_{k=1}^{K}\beta_{k}/(K\sigma_{w}^{2}))$
and 
\begin{align}
\Lambda_{\mathrm{{\scriptscriptstyle IGMM}}} & =2\,\frac{\bm{\|y}\|^{2}\,\sum_{k=1}^{K}\beta_{k}(P_{D,k}-P_{F,k})}{(\sum_{k=1}^{K}\beta_{k}P_{D,k}+\sigma_{w}^{2})(\sum_{k=1}^{K}\beta_{k}P_{F,k}+\sigma_{w}^{2})}.
\end{align}
The latter result follows from $\mathbb{E}\{\underline{\bm{y}}|\mathcal{H}_{i}\}=\bm{0}_{2N}$
and $\bm{\Sigma}_{\underline{\bm{y}}|\mathcal{H}_{1}}=(\sum_{k=1}^{K}\beta_{k}P_{i,k}+\sigma_{w}^{2})\,\bm{I}_{2N}$
(since under NLOS assumption $\bm{R}=\bm{O}_{K}$ implies $\widetilde{\bm{A}}(\bm{\theta})=\bm{O}_{N\times K}$
and $\nu_{k}=\beta_{k}$, respectively). Therefore it is apparent
that IS and IGMM (assuming $(P_{D,k}-P_{F.k})\geq0$) rules become
statistically equivalent to NLOS rule. 
\end{IEEEproof}
The above lemma states that IS and IGMM rules are both statistically
equivalent to NLOS rule when each sensor has only a purely scattered
component. Indeed, in such a case (as also supported intuitively),
only a\emph{ dependence} on $\left\Vert \bm{y}\right\Vert ^{2}$ is
relevant in the design of a fusion rule for the binary hypothesis
test under consideration (i.e., all the mentioned decision procedures
collapse into the received energy test). Accordingly IS rule, being
based on a weighted combination of a MRC-ED (see Eq. (\ref{eq: IS rule})),
exploits only the non-coherent term in NLOS case. Similarly IGMM rule,
being based on second-order characterization of $\bm{y}|\mathcal{H}_{i}$,
simplifies as the two hypotheses manifest in NLOS scenario with a
sole change of variance in the received signal (i.e., no mean or covariance
structure modification).

However, in the case of conditionally i.i.d. decisions, a \emph{stronger}
result can be proved for IS and IGMM rules, as described by the following
lemma.
\begin{lem}
\label{lem: Approx NLOS IGMM}In the case of conditionally i.i.d.
decisions, viz $P(\bm{x}|\mathcal{H}_{i})=\prod_{k=1}^{K}P(x_{k}|\mathcal{H}_{i})$
and $(P_{D,k},P_{F,k})=(P_{D},P_{F})$ (recalling that $P_{D}>P_{F}$),
and under a ``weak-LOS assumption'' (quantified as $P_{i}(1-P_{i})\lambda_{\mathrm{min}}(\underline{\widetilde{\bm{A}}}(\bm{\theta})\,\underline{\widetilde{\bm{A}}}(\bm{\theta})^{\dagger})\ll\sigma_{e,i}^{2}$),
IS and IGMM rules are approximately statistically equivalent.
\end{lem}
\begin{IEEEproof}
We begin by observing that, under conditionally i.i.d. assumption,
the covariance of $\underline{\bm{y}}|\mathcal{H}_{i}$ simplifies
to (since $\bm{\Sigma}_{\bm{x}|\mathcal{H}_{i}}=P_{i}(1-P_{i})\,\bm{I}_{K}$):
\begin{equation}
\bm{\Sigma}_{\underline{\bm{y}}|\mathcal{H}_{i}}=P_{i}(1-P_{i})\,\underline{\widetilde{\bm{A}}}(\bm{\theta})\,\underline{\widetilde{\bm{A}}}(\bm{\theta})^{\dagger}+\sigma_{e,i}^{2}\,\bm{I}_{2N}\,.
\end{equation}
Then, we express it in terms of the eigendecomposition $(\underline{\widetilde{\bm{A}}}(\bm{\theta})\,\underline{\widetilde{\bm{A}}}(\bm{\theta})^{\dagger})=(\bm{U}_{M}\,\bm{\Lambda}_{M}\,\bm{U}_{M}^{\dagger})$,
that is $\bm{\Sigma}_{\underline{\bm{y}}|\mathcal{H}_{i}}=\bm{U}_{M}\,[P_{i}(1-P_{i})\bm{\Lambda}_{M}+\sigma_{e,i}^{2}\bm{I}_{2N}]\,\bm{U}_{M}^{\dagger}$.
If $P_{i}(1-P_{i})\mbox{\ensuremath{\mathrm{eig}}}_{\mathrm{min}}(\underline{\widetilde{\bm{A}}}(\bm{\theta})\,\underline{\widetilde{\bm{A}}}(\bm{\theta})^{\dagger})\ll\sigma_{e,i}^{2}$
holds, we can safely approximate $\bm{\Sigma}_{\underline{\bm{y}}|\mathcal{H}_{i}}\approx(\bm{U}_{M}\sigma_{e,i}^{2}\,\bm{U}_{M}^{\dagger})$.
We refer to this assumption as a ``weak-LOS'' one since, as all
the $\kappa_{k}$'s get low in $\bm{R}$, all the eigenvalues in $\bm{\Lambda}_{M}$
get small while $\sigma_{e,i}^{2}=\left(P_{i}\,\sum_{k=1}^{K}\bm{D}(\bm{I}_{K}-\bm{R}^{2})\bm{1}_{K}+\sigma_{w}^{2}\right)$
increases. Also, we notice that IGMM rule in Eq. (\ref{eq: IGMM})
is statistically equivalent to:
\begin{gather}
\underline{\bm{y}}{}^{\dagger}\,(\bm{\Sigma}_{\underline{\bm{y}}|\mathcal{H}_{0}}^{-1}-\bm{\Sigma}_{\underline{\bm{y}}|\mathcal{H}_{1}}^{-1})\,\underline{\bm{y}}\,+\nonumber \\
2\,\underline{\bm{y}}^{\dagger}\left(\bm{\Sigma}_{\underline{\bm{y}}|\mathcal{H}_{1}}^{-1}\mathbb{E}\{\underline{\bm{y}}|\mathcal{H}_{1}\}-\bm{\Sigma}_{\underline{\bm{y}}|\mathcal{H}_{0}}^{-1}\mathbb{E}\{\underline{\bm{y}}|\mathcal{H}_{0}\}\right)\label{eq: IGMM_alt}
\end{gather}
Thus, by exploiting the aforementioned approximation in Eq.~(\ref{eq: IGMM_alt}),
$\Lambda_{\mathrm{{\scriptscriptstyle IGMM}}}$ is shown to be approximately
expressed as:
\begin{gather}
\bm{\|y}\|^{2}\,\left[\frac{1}{\sigma_{e,0}^{2}}-\frac{1}{\sigma_{e,1}^{2}}\right]+2\,\underline{\bm{y}}^{\dagger}\left[\frac{\mathbb{E}\{\underline{\bm{y}}|\mathcal{H}_{1}\}}{\sigma_{e,1}^{2}}-\frac{\mathbb{E}\{\underline{\bm{y}}|\mathcal{H}_{0}\}}{\sigma_{e,0}^{2}}\right]\label{eq: Approx IGMM}
\end{gather}
Then, after few manipulations (and exploiting definition of $\sigma_{e,i}^{2}$
and $\mathbb{E}\{\bm{y}|\mathcal{H}_{1}\}$, respectively), Eq. (\ref{eq: Approx IGMM})
can be rewritten as:
\begin{gather}
\frac{2\,(P_{D}-P_{F})}{\sigma_{e,0}^{2}\,\sigma_{e,1}^{2}}\,(\sum_{k=1}^{K}\nu_{k}\,\bm{\|y}\|^{2}+\sigma_{w}^{2}\,\underline{\bm{y}}^{\dagger}\underline{\widetilde{\bm{A}}}(\bm{\theta})\,\bm{1}_{K})=\nonumber \\
\frac{2K\,(P_{D}-P_{F})\,\sigma_{w}^{2}}{\sigma_{e,0}^{2}\,\sigma_{e,1}^{2}}\,\left(\frac{\bar{\nu}}{\sigma_{w}^{2}}\bm{\|y}\|^{2}+2\,\Re(\bar{\bm{\mu}}^{\dagger}\bm{y})\right)\label{eq: IGMM_lemma2_weak los}
\end{gather}
which is apparently the IS rule (except for an irrelevant positive
scalar, recalling $P_{D}>P_{F}$). This concludes the proof.
\end{IEEEproof}
We underline that Lem. \ref{lem: Exact NLOS} \emph{does not include}
Lem. \ref{lem: Approx NLOS IGMM}, since at a relatively low Rician
factor ($\mbox{\ensuremath{\mathrm{eig}}}_{\mathrm{min}}(\underline{\widetilde{\bm{A}}}(\bm{\theta})\,\underline{\widetilde{\bm{A}}}(\bm{\theta})^{\dagger})$
gets low, whereas $\sigma_{e,i}^{2}$ increases) for all the sensors,
data covariance matrix under $\mathcal{H}_{i}$ will be approximately
diagonal, while the difference of the mean terms $\mathbb{E}\{\underline{\bm{y}}|\mathcal{H}_{i}\}$
will not be negligible. In the latter case, IGMM will exhibit the
same linear-quadratic dependence on the data as the IS rule. In other
terms, it will reduce to a weighted combination of a MRC-ED (see (\ref{eq: IGMM_lemma2_weak los})
and (\ref{eq: IS rule}), respectively). In this region, however,
NLOS rule does not perform as well as those statistics, since its
dependence is \emph{only on} $\left\Vert \bm{y}\right\Vert ^{2}$.
Moreover, it is worth noticing that weak-LOS assumption $P_{i}(1-P_{i})\mathrm{eig}_{\mathrm{min}}(\underline{\widetilde{\bm{A}}}(\bm{\theta})\,\underline{\widetilde{\bm{A}}}(\bm{\theta})^{\dagger})\ll\sigma_{e,i}^{2}$
is also likely to be satisfied in a low-SNR regime (i.e., high $\sigma_{w}^{2}$,
right-hand increases) and for ``good-quality'' sensors (i.e., $(P_{D},P_{F})\rightarrow(1,0)$,
left-hand decreases).

Finally, we look at the extreme case given by IS assumption. In this
case, IS rule is statistically equivalent to the LLR (by construction,
cf. Sec. \ref{subsec: IS}). On the other hand, we are able to prove
the following asymptotic equivalence properties among WL and IGMM
rules, reported in the following lemma.
\begin{lem}
\label{lem: IS assumption}Under ``IS assumption'', IGMM rule is
statistically equivalent to IS rule (thus attaining optimum performance),
while WL rules are statistically equivalent and are given by the sole
``widely-linear'' part of IS rule in Eq.~(\ref{eq: IS rule}).
\end{lem}
\begin{IEEEproof}
We start by recalling statistical equivalence of IGMM rule to Eq.
(\ref{eq: IGMM_alt}). Then, we observe that IS assumption straightforwardly
implies $\bm{\rho}_{1}=\bm{1}_{K}$ (resp. $\bm{\rho}_{0}=\bm{0}_{K}$)
and $\bm{\Sigma}_{\bm{x}|\mathcal{H}_{i}}=\bm{O}_{K}$. Thus Eq. (\ref{eq: IGMM_alt})
specializes into:
\begin{gather}
\underline{\bm{y}}{}^{\dagger}\,\left(\frac{\sigma_{e,1}^{2}-\sigma_{e,0}^{2}}{\sigma_{e,0}^{2}\,\sigma_{e,1}^{2}}\right)\,\underline{\bm{y}}+\frac{2}{\sigma_{e,1}^{2}}\,\underline{\bm{y}}^{\dagger}\underline{\widetilde{\bm{A}}}(\bm{\theta})\,\bm{1}_{K}=\nonumber \\
\frac{2K}{\sigma_{e,1}^{2}}\,\left\{ \left\Vert \bm{y}\right\Vert ^{2}\,(\bar{\nu}/\sigma_{e,0}^{2})+2\,\Re\{\bar{\bm{\mu}}^{\dagger}\bm{y}\}\right\} \label{eq: IGMM_IS specialized}
\end{gather}
which is related to IS rule via an irrelevant positive constant (we
recall that, under IS assumption, $\sigma_{e,1}^{2}=(\sum_{k=1}^{K}\nu_{k}+\sigma_{w}^{2})$
and $\sigma_{e,0}^{2}=\sigma_{w}^{2}$ hold). This proves the first
part of the lemma. By similar reasoning, it can be shown that both
WL rules in Eq.~(\ref{eq: Deflection maximizer WL formula}), under
IS assumption, coincide with:
\begin{equation}
\underline{\bm{y}}^{\dagger}\left((\underline{\widetilde{\bm{A}}}(\bm{\theta})\,\bm{1}_{K})/\left\Vert \underline{\widetilde{\bm{A}}}(\bm{\theta})\,\bm{1}_{K}\right\Vert \right)=\frac{\sqrt{2}}{\left\Vert \bar{\bm{\mu}}\right\Vert }\,\Re\{\bar{\bm{\mu}}^{\dagger}\bm{y}\}\,.\label{eq: WL_IS specialized}
\end{equation}
It is apparent that right-hand side of Eq.~(\ref{eq: WL_IS specialized})
is proportional to the first contribution of IS rule in Eq. (\ref{eq: IS rule}),
thus completing the proof.
\end{IEEEproof}
Therefore, when sensors are ideal, IGMM rule will be statistically
equivalent to IS (viz. LLR) rule, as no covariance structure change
happens when one of the two hypotheses is in force. Differently WL
rules, lacking a $\left\Vert \bm{y}\right\Vert ^{2}$ dependence,
do not reduce to a weighted MRC-ED combination. Based on this reason,
we expect that when the WSN operates with ``good-quality'' sensors,
WL rules will experience some performance loss with respect to IS
and IGMM rules.

\section{Jammer (Subspace) Interference Environment\label{sec: Subspace Interference}}

In this section, we complicate the model in Eq. (\ref{eq:Channel_model})
and assume the presence of jamming devices operating on the WSN-DFC
communication channel. More specifically, we model the jamming signal
as an $r$-dimensional vector, whose experienced channel follows the
same Rician model as the WSN at the DFC, that is $\bm{y}_{s}=\bm{y}+\bm{s}_{J}$,
where: 
\begin{align}
\bm{s}_{J} & =\left(\bm{A}_{J}(\bm{\phi})\,\bm{R}_{J}+\bm{H}_{J}\,(\bm{I}_{r}-\bm{R}_{J}^{2})^{1/2}\right)\,\bm{D}_{J}^{1/2}\bm{\psi}.\label{eq: additional interference signal}
\end{align}
 In Eq. (\ref{eq: additional interference signal}) $\bm{\psi}\in\mathbb{C}^{r}$
represents the (unknown deterministic) jamming signal. Similarly to
the WSN, $\bm{A}_{J}(\bm{\phi})\in\mathbb{C}^{N\times r}$, $\bm{H}_{J}\in\mathbb{C}^{N\times r}$,
$\bm{R}_{J}\in\mathbb{R}^{r\times r}$ and $\bm{D}_{J}\in\mathbb{R}^{r\times r}$
denote the (full-rank) steering matrix (whose $\ell$th column is
given by $\bm{a}(\phi_{\ell})$ and depends on the angle-of-arrival
$\phi_{\ell}$), the normalized scattered matrix (whose $\ell$th
column $\bm{h}_{J,\ell}\sim\mathcal{N}_{\mathbb{C}}(\bm{0}_{N},\bm{I}_{N})$,
assumed mutually independent from the others), the diagonal matrix
of the Rician factors (whose $\ell$th element is denoted as $b_{\ell,J}$)
and the large-scale diagonal fading matrix of the jammer (whose $\ell$th
element is denoted as $\beta_{\ell,J}$), respectively. It is worth
noticing that Eq.~(\ref{eq: additional interference signal}) accounts
for interfering systems with both distributed (viz. $\bm{R}_{J}$
and $\bm{D}_{J}$ are both diagonal) or co-located (viz. $\bm{R}_{J}$
and $\bm{D}_{J}$ are both scaled identity) transmitting antennas
in space \cite{Gesbert2002}. It is apparent that the former case
includes the case of \emph{multiple} \emph{jammers}. The considered
interfering source can be classified as a ``constant jammer'', according
to the terminology\footnote{We underline that the term ``constant'' may be misleading, as the
definition of \cite{Xu2006} implies that the jammer continuously
emits a radio signal (changing with time), which is unknown at the
DFC.} proposed in \cite{Xu2006}. Though it represents the simplest typology
of jammer, it is here considered as a first step toward the development
of fusion rules robust to ``smarter'' jammers.

In this case, the received signal $\bm{y}_{s}$ is conditionally distributed
as:
\begin{gather}
\bm{y}_{s}|\mathcal{H}_{i}\sim\sum_{\bm{x}\in\mathcal{X}^{K}}P(\bm{x}|\mathcal{H}_{i})\,\mathcal{N}_{\mathbb{C}}\left(\bm{\mu}_{s}(\bm{x},\bm{\zeta}),[\sigma_{e}^{2}(\bm{x})+\sigma_{J}^{2}]\,\bm{I}_{N}\right)\nonumber \\
\mathrm{where}\quad\quad\bm{\mu}_{s}(\bm{x},\bm{\zeta})\triangleq\widetilde{\bm{A}}(\bm{\theta})\,\bm{x}+\bm{A}_{J}(\bm{\phi})\,\bm{\zeta}
\end{gather}
where $\sigma_{J}^{2}\triangleq\sum_{\ell=1}^{r}\nu_{\ell,J}\,|\psi_{\ell}|^{2}$,
$\nu_{\ell,J}\triangleq\beta_{\ell,J}\,(1-b_{\ell,J}^{2})$ and $\bm{\zeta}\triangleq(\bm{R}_{J}\,\bm{D}_{J}^{1/2}\,\bm{\psi})$,
respectively. Hereinafter we will make the reasonable assumption that
the DFC can only \emph{learn} $\bm{A}_{J}(\bm{\phi})$, i.e., the
DFC does not have knowledge of: ($i$) the diagonal matrix of the
Rician factors $\bm{R}_{J}$, ($ii$) the large-scale fading diagonal
matrix $\bm{D}_{J}$ and ($iii)$ the actual jamming (transmitted)
signal $\bm{\psi}$. The following sub-sections are thus devoted to
the design of (sub-optimal) fusion rules in the presence of the aforementioned
(unknown deterministic) interference parameters.

\subsection{Clairvoyant LRT and GLRT }

In what follows, we will employ in our comparison the clairvoyant
LRT as a benchmark, which (unrealistically) assumes $\{\bm{\psi},\bm{D}_{J},\bm{R}_{J}\}$
as known and thus implements the statistic:
\begin{gather}
\Lambda_{\mathrm{c-opt}}\triangleq\label{eq: Clairvoyant LRT rule}\\
\ln\left[\frac{\sum_{\bm{x}\in\mathcal{X}^{K}}\frac{P(\bm{x}|\mathcal{H}_{1})}{[\sigma_{e}^{2}(\bm{x})+\sigma_{J}^{2}]^{N}}\exp(-\frac{\bm{\|y}-\sum_{k=1}^{K}\bm{\mu}_{k}\,x_{k}-\bm{A}_{J}(\bm{\phi})\,\bm{\zeta}\|^{2}}{\sigma_{e}^{2}(\bm{x})+\sigma_{J}^{2}})}{\sum_{\bm{x}\in\mathcal{X}^{K}}\frac{P(\bm{x}|\mathcal{H}_{0})}{[\sigma_{e}^{2}(\bm{x})+\sigma_{J}^{2}]^{N}}\exp(-\frac{\bm{\|y}-\sum_{k=1}^{K}\bm{\mu}_{k}\,x_{k}-\bm{A}_{J}(\bm{\phi})\,\bm{\zeta}\|^{2}}{\sigma_{e}^{2}(\bm{x})+\sigma_{J}^{2}})}\right]\nonumber 
\end{gather}
Clearly, the LRT is \emph{uniformly most powerful }\cite{Lehmann2006}
and thus no other fusion rule can expect to perform better. Unfortunately
the LRT cannot be implemented, as the jamming parameters are not known
in the practice. For this reason, hereinafter we will devise tests
which tackle the arising composite hypothesis testing problem. 

A widespread test for the considered problem would be the GLRT \cite{Kay1998},
requiring the maximization of pdf under both hypotheses w.r.t. the
(unknown) parameters set. The GLRT has been successfully applied to
different application contexts, such as spectrum sensing \cite{Zhang2010},
allowing important design guidelines on system level performance (in
terms of optimized sensing time) \cite{He2016}. In our case, it is
not difficult to show that optimization w.r.t. $\{\bm{\psi},\bm{D}_{J},\bm{R}_{J}\}$
is tantamount to maximizing both pdfs w.r.t. $\sigma_{J}^{2}$ and
$\bm{\zeta}$ as they were \emph{(parametrically) independent}. Therefore,
this yields the statistic:
\begin{equation}
\Lambda_{\mathrm{{\scriptscriptstyle GLR}}}\triangleq\ln\left[\frac{\max_{\bm{\zeta},\sigma_{J}^{2}}\,p(\bm{y}_{s}|\mathcal{H}_{1})}{\max_{\bm{\zeta},\sigma_{J}^{2}}\,p(\bm{y}_{s}|\mathcal{H}_{0})}\right]\,.\label{eq: GLRT exploited}
\end{equation}
From inspection of Eq. (\ref{eq: GLRT exploited}), it is apparent
that GLRT has no simple implementation for this problem, because of
its exponential complexity ($p(\bm{y}_{s}|\mathcal{H}_{i})$ is a
GM with $2^{K}$ components) and required non-linear optimizations.
Thus, exact GLRT implementation appears as not feasible from a practical
point of view and will not be pursued in the following. Nonetheless,
we will show that ``GLRT philosophy'' of Eq.~(\ref{eq: GLRT exploited})
can be exploited jointly with the simplifying assumptions that lead
to the sub-optimal statistics obtained in Sec. \ref{sec:Fusion-Rules}
in order to devise \emph{computationally efficient} and \emph{jamming-robust}
fusion rules.

\subsection{IS-GLRT rule}

The GLRT in Eq.~(\ref{eq: GLRT exploited}) can be simplified under
the IS assumption, i.e., $P(\bm{x}=\bm{1}_{K}|\mathcal{H}_{1})=P(\bm{x}=\bm{0}_{K}|\mathcal{H}_{0})=1$.
Indeed, based on these assumptions, it holds:
\begin{align}
\bm{y}_{s}|\mathcal{H}_{0} & \sim\mathcal{N}_{\mathbb{C}}(\bm{A}_{J}(\bm{\phi})\,\bm{\zeta},\,[\sigma_{w}^{2}+\sigma_{J}^{2}]\,\bm{I}_{N})\label{eq: Char- IS GLRT}\\
\bm{y}_{s}|\mathcal{H}_{1} & \sim\mathcal{N}_{\mathbb{C}}(\widetilde{\bm{A}}(\bm{\theta})\,\bm{1}_{K}+\bm{A}_{J}(\bm{\phi})\,\bm{\zeta},\,[\sigma_{e}^{2}(\bm{1}_{K})+\sigma_{J}^{2}]\,\bm{I}_{N})\nonumber 
\end{align}
The ML estimates of $\bm{\zeta}$ under $\mathcal{H}_{0}$ and $\mathcal{H}_{1}$
are obtained respectively as \cite{Kay1998}:
\begin{align}
\hat{\bm{\zeta}}_{0}\triangleq\, & \bm{A}_{J}(\bm{\phi})^{-}\,\bm{y}_{s}\label{eq: ML estimate zita (H0) IS GLRT}\\
\hat{\bm{\zeta}}_{1}\triangleq\, & \bm{A}_{J}(\bm{\phi})^{-}\,(\bm{y}_{s}-\widetilde{\bm{A}}(\bm{\theta})\,\bm{1}_{K})\label{eq: ML estimate zita (H1)  IS GLRT}
\end{align}
Hence, the concentrated likelihoods are:
\begin{gather}
p_{\mathrm{is}}(\bm{y}_{s}|\mathcal{H}_{0},\hat{\bm{\zeta}}_{0},\sigma_{J}^{2})=\nonumber \\
\frac{1}{\{\pi[\sigma_{w}^{2}+\sigma_{J}^{2}]\}^{N}}\,\exp\left[-\frac{\left\Vert \bm{r}_{0}\right\Vert ^{2}}{\sigma_{w}^{2}+\sigma_{J}^{2}}\right]\label{eq:conc_likelihood_H0 IS GLRT}\\
p_{\mathrm{is}}(\bm{y}_{s}|\mathcal{H}_{1},\hat{\bm{\zeta}}_{1},\sigma_{J}^{2})=\nonumber \\
\frac{1}{\{\pi[\sigma_{e}^{2}(\bm{1}_{K})+\sigma_{J}^{2}]\}^{N}}\,\exp\left[-\frac{\left\Vert \bm{r}_{1}\right\Vert ^{2}}{\sigma_{e}^{2}(\bm{1}_{K})+\sigma_{J}^{2}}\right]\label{eq:conc_likelihood_H1  IS GLRT}
\end{gather}
where $\bm{r}_{0}\triangleq[\bm{P}_{\bm{A}_{J}(\bm{\phi})}^{\perp}\,\bm{y}_{s}]$
and $\bm{r}_{1}\triangleq[\bm{P}_{\bm{A}_{J}(\bm{\phi})}^{\perp}(\bm{y}_{s}-\widetilde{\bm{A}}(\bm{\theta})\,\bm{1}_{K})]$,
respectively. Then the ML estimates\footnote{These are straightforwardly obtained by setting $\frac{\partial\ln\,p_{\mathrm{is}}(\bm{y}_{s}|\mathcal{H}_{i},\hat{\bm{\zeta}}_{i},\sigma_{J}^{2})}{\partial\sigma_{J}^{2}}=0$
and accounting for the constraint $\sigma_{J}^{2}\geq0$.} of $\sigma_{J}^{2}$ under $\mathcal{H}_{0}$ and $\mathcal{H}_{1}$
are obtained as \cite{Kay1998}:
\begin{eqnarray}
\hat{\sigma}_{J,0}^{2} & \triangleq & \left[\nicefrac{\left\Vert \bm{r}_{0}\right\Vert ^{2}}{N}-\sigma_{w}^{2}\right]_{+}\label{eq: IS-GLRT - sigmaj estimate H0}\\
\hat{\sigma}_{J,1}^{2} & \triangleq & \left[\nicefrac{\left\Vert \bm{r}_{1}\right\Vert ^{2}}{N}-\sigma_{e}^{2}(\bm{1}_{K})\right]_{+}\label{eq: IS-GLRT - sigmaj estimate H1}
\end{eqnarray}
Then, we substitute Eqs. (\ref{eq: IS-GLRT - sigmaj estimate H0})
and (\ref{eq: IS-GLRT - sigmaj estimate H1}) into (\ref{eq:conc_likelihood_H0 IS GLRT})
and (\ref{eq:conc_likelihood_H1  IS GLRT}), respectively, thus obtaining:
\begin{gather}
p_{\mathrm{is}}(\bm{y}_{s}|\mathcal{H}_{0},\hat{\bm{\zeta}}_{0},\hat{\sigma}_{J,0}^{2})=\nonumber \\
\frac{1}{\{\pi[\sigma_{w}^{2}+\hat{\sigma}_{J,0}^{2}]\}^{N}}\,\exp\left[-\frac{\left\Vert \bm{r}_{0}\right\Vert ^{2}}{\sigma_{w}^{2}+\hat{\sigma}_{J,0}^{2}}\right]\\
p_{\mathrm{is}}(\bm{y}_{s}|\mathcal{H}_{1},\hat{\bm{\zeta}}_{1},\hat{\sigma}_{J,1}^{2})=\nonumber \\
\frac{1}{\{\pi[\sigma_{e}^{2}(\bm{1}_{K})+\hat{\sigma}_{J,1}^{2}]\}^{N}}\,\exp\left[-\frac{\left\Vert \bm{r}_{1}\right\Vert ^{2}}{\sigma_{e}^{2}(\bm{1}_{K})+\hat{\sigma}_{J,1}^{2}}\right]
\end{gather}
Taking $\ln(\cdot)$ of the concentrated likelihood ratio $\left\{ p_{\mathrm{is}}(\bm{y}_{s}|\mathcal{H}_{1},\hat{\bm{\zeta}}_{1},\hat{\sigma}_{J,1}^{2})/p_{\mathrm{is}}(\bm{y}_{s}|\mathcal{H}_{0},\hat{\bm{\zeta}}_{0},\hat{\sigma}_{J,0}^{2})\right\} $
provides the final expression:
\begin{align}
\Lambda_{{\scriptscriptstyle \mathrm{IS-GLR}}}\triangleq & \left\{ \,N\,\ln\left[\frac{\sigma_{w}^{2}+\hat{\sigma}_{J,0}^{2}}{\sigma_{e}^{2}(\bm{1}_{K})+\hat{\sigma}_{J,1}^{2}}\right]-\frac{\left\Vert \bm{r}_{1}\right\Vert ^{2}}{\sigma_{e}^{2}(\bm{1}_{K})+\hat{\sigma}_{J,1}^{2}}\right.\nonumber \\
 & \left.+\frac{\left\Vert \bm{r}_{0}\right\Vert ^{2}}{\sigma_{w}^{2}+\hat{\sigma}_{J,0}^{2}}\right\} \label{eq: IS-GLRT}
\end{align}
The proposed rule, in analogy to Sec. \ref{subsec: IS}, will be referred
to as \emph{IS-GLRT} in the following.

\subsection{NLOS-GLRT rule}

Differently, here we start by using the NLOS assumption ($\kappa_{k}=0$,
$k\in\mathcal{K}$) on the conditional received signal pdf, which
gives:
\begin{gather}
\bm{y}_{s}|\mathcal{H}_{i}\sim\sum_{\bm{x}\in\mathcal{X}^{K}}P(\bm{x}|\mathcal{H}_{i})\,\mathcal{N}_{\mathbb{C}}(\bm{A}_{J}(\bm{\phi})\,\bm{\zeta},\,[\bar{\sigma}_{e}^{2}(\bm{x})+\sigma_{J}^{2}]\,\bm{I}_{N})\label{eq: NLOS-GLRT signal model}
\end{gather}
where $\bar{\sigma}_{e}^{2}(\bm{x})\triangleq(\sigma_{w}^{2}+\sum_{k=1}^{K}\beta_{k}x_{k})$.
Even under such a simplifying assumption, Eq. (\ref{eq: NLOS-GLRT signal model})
still has the form of a complex Gaussian mixture with $2^{K}$ distinct
components, thus being intractable from a computational point of view.
Thus, we further resort to Gaussian moment matching to fit the pdf
of $\bm{y}_{s}|\mathcal{H}_{i}$ to a (proper) complex Gaussian pdf
as follows:
\begin{gather}
\mathbb{E}\{\bm{y}_{s}|\mathcal{H}_{i}\}=\bm{A}_{J}(\bm{\phi})\,\bm{\zeta}\qquad\bm{\Sigma}_{\bm{y}_{s}|\mathcal{H}_{i}}=(\sigma_{n,i}^{2}+\sigma_{J}^{2})\,\bm{I}_{N}
\end{gather}
where we have denoted $\sigma_{n,i}^{2}\triangleq(\sum_{k=1}^{K}P_{i,k}\,\beta_{k}+\sigma_{w}^{2})$.
Therefore, moment matching yields:
\begin{equation}
\bm{y}_{s}|\mathcal{H}_{i}\sim\mathcal{N}_{\mathbb{C}}(\bm{A}_{J}(\bm{\phi})\,\bm{\zeta},\,[\sigma_{n,i}^{2}+\sigma_{J}^{2}]\,\bm{I}_{N})\label{eq: matched_distribution}
\end{equation}
Now, in order to obtain a GLRT-like statistic, we would need to evaluate
the ML estimates of $\{\bm{\zeta},\sigma_{J}^{2}\}$ under $\mathcal{H}_{i}$
for the matched model in Eq. (\ref{eq: matched_distribution}). This
is the case for the ML estimates of $\bm{\zeta}$ under $\mathcal{H}_{0}$
and $\mathcal{H}_{1}$, being \emph{both} \emph{equal} to $\hat{\bm{\zeta}}_{0}$
(cf. Eq.~(\ref{eq: ML estimate zita (H0) IS GLRT})). After substitution,
the concentrated matched likelihood of $\bm{y}_{s}|\mathcal{H}_{i}$
is:
\begin{gather}
p_{\mathrm{nl}}(\bm{y}_{s}|\mathcal{H}_{i};\hat{\bm{\zeta}},\sigma_{J}^{2})=\frac{1}{\{\pi[\sigma_{n,i}^{2}+\sigma_{J}^{2}]\}^{N}}\,\exp\left[-\frac{\left\Vert \bm{r}_{0}\right\Vert ^{2}}{\sigma_{n,i}^{2}+\sigma_{J}^{2}}\right]\label{eq: conc_likelihood_H0-1-1-1}
\end{gather}
where $\bm{r}_{0}$ has the same meaning as for IS-GLRT rule. After
substitution, it is not difficult to prove that the ``moment-matched''
concentrated likelihood ratio
\begin{equation}
\frac{p_{\mathrm{nl}}(\bm{y}_{s}|\mathcal{H}_{1};\hat{\bm{\zeta}}_{0},\sigma_{J}^{2})}{p_{\mathrm{nl}}(\bm{y}_{s}|\mathcal{H}_{0};\hat{\bm{\zeta}}_{0},\sigma_{J}^{2})}\label{eq: Matched_LLR_NLOS_GLRT}
\end{equation}
is an increasing function of $\left\Vert \bm{r}_{0}\right\Vert ^{2}$,
independently on the value of the (unknown) $\sigma_{J}^{2}$, whose
estimation can be thus avoided (the proof can be obtained by taking
the logarithm of (\ref{eq: Matched_LLR_NLOS_GLRT}) and exploiting
$P_{D,k}\geq P_{F,k}$). Therefore, the test deciding for $\mathcal{H}_{1}$
when $\Lambda_{{\scriptscriptstyle \mathrm{NL-GLR}}}>\gamma$, where
\begin{equation}
\Lambda_{{\scriptscriptstyle \mathrm{NL-GLR}}}\triangleq\left\Vert \bm{P}_{\bm{A}_{J}(\bm{\phi})}^{\perp}\bm{y}_{s}\right\Vert ^{2}\label{eq: NLOS-GLRT}
\end{equation}
is uniformly most powerful under NLOS assumption and after moment
matching. For the mentioned reason, the present test, denoted here
as \emph{NLOS-GLRT} (in analogy to Sec. \ref{subsec: NLOS rule} and
with a slight abuse of terminology, since estimation of $\sigma_{J}^{2}$
is not needed for test implementation), will be employed in our comparison.

\subsection{IGMM-GLRT rule\label{subsec:IGMM-GLRT-rule}}

It can be readily shown that the characterization up to the second
order in Eqs. (\ref{eq: 2nd order char MEAN}) and (\ref{eq: 2nd order char PCOV})
generalizes to:
\begin{gather}
\mathbb{E}\{\bm{y}_{s}|\mathcal{H}_{i}\}=\bm{t}_{i}+\bm{A}_{J}(\bm{\phi})\,\bm{\zeta}\label{eq: IGMM-GLRT char 1}\\
\bm{\Sigma}_{\bm{y}_{s}|\mathcal{H}_{i}}=\bm{\Sigma}_{\bm{y}|\mathcal{H}_{i}}+\sigma_{J}^{2}\,\bm{I}_{N}\qquad\bar{\bm{\Sigma}}_{\bm{y}_{s}|\mathcal{H}_{i}}=\bar{\bm{\Sigma}}_{\bm{y}|\mathcal{H}_{i}}\label{eq: IGMM-GLRT char 2}
\end{gather}
where we have denoted $\bm{t}_{i}\triangleq\mathbb{E}\{\bm{y}|\mathcal{H}_{i}\}$
(cf. Eq. (\ref{eq: 2nd order char MEAN})). We first match the pdf
of the vector $\bm{y}_{s}|\mathcal{H}_{i}$ to that of an improper
complex Gaussian vector, that is:
\begin{equation}
\bm{y}_{s}|\mathcal{H}_{i}\sim\mathcal{N}_{\mathbb{C}}(\bm{t}_{i}+\bm{A}_{J}(\bm{\phi})\,\bm{\zeta},\bm{\Sigma}_{\bm{y}_{s}|\mathcal{H}_{i}},\bar{\bm{\Sigma}}_{\bm{y}_{s}|\mathcal{H}_{i}})\label{eq: Improper Gaussian moment matching}
\end{equation}
It is easy to verify that Eq. (\ref{eq: Improper Gaussian moment matching})
is also equivalent to the following linear model:
\begin{equation}
\bm{y}_{s}|\mathcal{H}_{i}=\bm{t}_{i}+\bm{A}_{J}(\bm{\phi})\,\bm{\zeta}+\bm{w}_{i}
\end{equation}
where $\bm{w}_{i}\sim\mathcal{N}_{\mathbb{C}}(\bm{0}_{N},\bm{\Sigma}_{\bm{y}|\mathcal{H}_{i}}+\sigma_{J}^{2}\,\bm{I}_{N},\bar{\bm{\Sigma}}_{\bm{y}|\mathcal{H}_{i}})$
(i.e., a zero-mean non-circular complex Gaussian vector). Therefore,
when the hypothesis $\mathcal{H}_{i}$ is in force, we define $\bm{y}_{s,i}\triangleq(\bm{y}_{s}-\bm{t}_{i})$
and exploit the SVD of $\bm{A}_{J}(\bm{\phi})=(\bm{U}_{J}\,\bm{\Lambda}_{J}\,\bm{V}_{J}^{\dagger})$,
thus obtaining 
\begin{equation}
\bm{y}_{s,i}=\bm{U}_{J}\,\underbrace{\begin{bmatrix}\bm{\Lambda}_{r}\\
\bm{O}_{(N-r)\times r}
\end{bmatrix}}_{=\,\bm{\Lambda}_{J}}\,\bm{V}_{J}^{\dagger}\,\bm{\zeta}+\bm{w}_{i}\,.\label{eq: WL-SVD decomposition-1}
\end{equation}
 where $\bm{\Lambda}_{r}\in\mathbb{C}^{r\times r}$ denotes the (diagonal)
sub-matrix extracted from the matrix of the singular values $\bm{\Lambda}_{J}$
(since the rank of the interference is equal to $r$). We then define
$\bm{\zeta}^{'}\triangleq(\bm{\Lambda}_{r}\,\bm{V}_{J}^{\dagger}\,\bm{\zeta})\in\mathbb{C}^{r}$
and notice that $\bm{\zeta}$ and $\bm{\zeta}^{'}$ are in \emph{one-to-one}
correspondence. Therefore, after a left-multiplication by $\bm{U}_{J}^{\dagger}$
(i.e., a unitary matrix, which does not entail loss of information),
Eq. (\ref{eq: WL-SVD decomposition-1}) is rewritten as follows:
\begin{align}
\bm{s}_{i} & =\underbrace{\begin{bmatrix}\bm{I}_{r}\\
\bm{O}_{(N-r)\times r}
\end{bmatrix}}_{\triangleq\bm{S}}\bm{\zeta}^{'}+\bm{n}_{i}
\end{align}
where $\bm{s}_{i}\triangleq(\bm{U}_{J}^{\dagger}\,\bm{y}_{s,i})\in\mathbb{C}^{N}$
and 
\begin{gather}
\bm{n}_{i}\sim\mathcal{N}_{\mathbb{C}}\left(\bm{0}_{N},\,\bm{U}_{J}^{\dagger}\,\bm{\Sigma}_{\bm{y}|\mathcal{H}_{i}}\,\bm{U}_{J}+\sigma_{J}^{2}\,\bm{I}_{N},\,\bm{U}_{J}^{\dagger}\,\bar{\bm{\Sigma}}_{\bm{y}|\mathcal{H}_{i}}\,\bm{U}_{J}^{*}\right)
\end{gather}
Then, we can define the following augmented model:
\begin{eqnarray}
\underline{\bm{s}_{i}} & = & \bar{\bm{S}}\,\underline{\bm{\zeta}^{'}}+\underline{\bm{n}_{i}}\\
\bar{\bm{S}} & \triangleq & \left[\begin{array}{cc}
\bm{I}_{r} & \bm{O}_{r}\\
\bm{O}_{(N-r)\times r} & \bm{O}_{(N-r)\times r}\\
\bm{O}_{r} & \bm{I}_{r}\\
\bm{O}_{(N-r)\times r} & \bm{O}_{(N-r)\times r}
\end{array}\right]
\end{eqnarray}
where $\underline{\bm{n}_{i}}\sim\mathcal{N}_{\mathbb{C}}(\bm{0}_{2N},\bm{R}_{A,i})$,
and we have defined $\bm{R}_{A,i}\triangleq(\bm{\Sigma}_{A,i}+\sigma_{J}^{2}\,\bm{I}_{2N})$,
$\bm{\Sigma}_{A,i}\triangleq\bar{\bm{U}}_{J}^{\dagger}\,\bm{\Sigma}_{\underline{\bm{y}}|\mathcal{H}_{i}}\,\bar{\bm{U}}_{J}$
and 
\begin{equation}
\bar{\bm{U}}_{J}\triangleq\begin{bmatrix}\bm{U}_{J} & \bm{O}_{N}\\
\bm{O}_{N} & \bm{U}_{J}^{*}
\end{bmatrix}\,.
\end{equation}
Hence, the (matched) pdf of the augmented vector $\underline{\bm{s}_{i}}$
is given by \cite{Schreier2010}:
\begin{gather}
p_{\mathrm{igmm}}\left(\underline{\bm{s}_{i}}\,;\underline{\bm{\zeta}^{'}},\sigma_{J}^{2}|\mathcal{H}_{i}\right)=\label{eq: WL transformed pdf-1}\\
\frac{1}{\pi^{N}\det(\bm{R}_{A,i})^{1/2}}\exp\left[-\frac{1}{2}\left(\underline{\bm{s}_{i}}-\bar{\bm{S}}\,\underline{\bm{\zeta}^{'}}\right)^{\dagger}\,\bm{R}_{A,i}^{-1}\,\left(\underline{\bm{s}_{i}}-\bar{\bm{S}}\,\underline{\bm{\zeta}^{'}}\right)\right]\nonumber 
\end{gather}
In order to obtain the IGMM-GLRT rule, we need the ML estimates of
$\{\underline{\bm{\zeta}^{'}},\sigma_{J}^{2}\}$. First, the ML estimate
of $\underline{\bm{\zeta}^{'}}$ from Eq.~(\ref{eq: WL transformed pdf-1})
is readily given by $\hat{\underline{\bm{\zeta}_{i}^{'}}}\triangleq(\bar{\bm{S}}{}^{\dagger}\,\bm{R}_{A,i}^{-1}\,\bar{\bm{S}})^{-1}\,\bar{\bm{S}}^{\dagger}\,\bm{R}_{A,i}^{-1}\,\underline{\bm{s}_{i}}$.
After substitution, the concentrated log-likelihood is:
\begin{gather}
\ln\,p_{\mathrm{igmm}}\left(\underline{\bm{s}_{i}};\hat{\underline{\bm{\zeta}_{i}^{'}}},\sigma_{J}^{2}|\mathcal{H}_{i}\right)=-N\,\ln\pi-\frac{1}{2}\ln\det(\bm{R}_{A,i})\nonumber \\
-\frac{1}{2}\underline{\bm{s}_{i}}^{\dagger}\,[\bm{R}_{A,i}^{-1}-\bm{R}_{A,i}^{-1}\,\bar{\bm{S}}\,(\bar{\bm{S}}^{\dagger}\,\bm{R}_{A,i}^{-1}\,\bar{\bm{S}})^{-1}\bar{\bm{S}}^{\dagger}\,\bm{R}_{A,i}^{-1}\,]\,\underline{\bm{s}_{i}}\label{eq: concentrated augmented likelihood}
\end{gather}
We now observe that $\bar{\bm{S}}$ is related to a conveniently defined
matrix $\bm{T}$ via a permutation matrix $\bm{\Gamma}$, as shown
in Eq. (\ref{eq: Perm matrix}) at the top of next page. 
\begin{figure*}
\begin{eqnarray}
\bar{\bm{S}} & = & \underbrace{\begin{bmatrix}\bm{I}_{r} & \bm{O}_{r} & \bm{O}_{r\times(N-r)} & \bm{O}_{r\times(N-r)}\\
\bm{O}_{(N-r)\times r} & \bm{O}_{(N-r)\times r} & \bm{I}_{(N-r)} & \bm{O}_{(N-r)}\\
\bm{O}_{r} & \bm{I}_{r} & \bm{O}_{r\times(N-r)} & \bm{O}_{r\times(N-r)}\\
\bm{O}_{(N-r)\times r} & \bm{O}_{(N-r)\times r} & \bm{O}_{(N-r)} & \bm{I}_{(N-r)}
\end{bmatrix}}_{\triangleq\,\bm{\Gamma}}\times\underbrace{\begin{bmatrix}\bm{I}_{2r}\\
\bm{O}_{2(N-r)\times2r}
\end{bmatrix}}_{\triangleq\,\bm{T}}\label{eq: Perm matrix}
\end{eqnarray}
\hrulefill 
\vspace*{0pt}
\end{figure*}
 Based on the aforementioned definition, Eq. (\ref{eq: concentrated augmented likelihood})
is rewritten as:
\begin{gather}
\ln\,p_{\mathrm{igmm}}\left(\underline{\bm{s}_{i}};\hat{\underline{\bm{\zeta}_{i}^{'}}},\sigma_{J}^{2}|\mathcal{H}_{i}\right)=-N\,\ln\pi-\frac{1}{2}\ln\det(\bm{R}_{A,i})\nonumber \\
-\frac{1}{2}\bm{m}_{i}^{\dagger}\,[\bm{R}_{p,i}^{-1}-\bm{R}_{p,i}^{-1}\,\bm{T}\,(\bm{T}^{\dagger}\,\bm{R}_{p,i}^{-1}\,\bm{T})^{-1}\bm{T}^{\dagger}\,\bm{R}_{p,i}^{-1}\,]\,\bm{m}_{i}\label{eq: concentrated augmented likelihood perm}
\end{gather}
where $\bm{m}_{i}\triangleq(\bm{\Gamma}^{\dagger}\,\underline{\bm{s}_{i}})$
and $\bm{R}_{p,i}\triangleq(\bm{\Gamma}^{\dagger}\,\bm{R}_{A,i}^{-1}\,\bm{\Gamma})^{-1}=(\bm{\Gamma}^{\dagger}\,\bm{R}_{A,i}\,\bm{\Gamma})$
(since every permutation matrix is unitary, i.e., $(\bm{\Gamma}^{\dagger}\bm{\Gamma})=(\bm{\Gamma}\bm{\Gamma}^{\dagger})=\bm{I}_{2N}$).
It can be recognized in second line of Eq.~(\ref{eq: concentrated augmented likelihood perm})
that matrix in square brackets has the block structure (obtained by
exploiting the simplified structure of $\bm{T}$) 
\begin{equation}
\begin{bmatrix}\bm{O}_{2r} & \bm{O}_{2r\times2(N-r)}\\
\bm{O}_{2(N-r)\times2r} & \bm{R}_{c,i}^{-1}
\end{bmatrix}
\end{equation}
 where $\bm{R}_{c,i}^{-1}\in\mathbb{C}^{2(N-r)\times2(N-r)}$ is the
Schur complement of the block $(\bm{R}_{p,i}^{-1})_{1:2r}$ of matrix
$\bm{R}_{p,i}^{-1}$ and can be identified from $\bm{R}_{p,i}$ as:
\begin{equation}
\bm{R}_{p,i}=\begin{bmatrix}\bm{\varTheta}_{i} & \bm{\varOmega}_{i}\\
\bm{\varOmega}_{i}^{\dagger} & \bm{R}_{c,i}
\end{bmatrix}
\end{equation}
where $\bm{\varTheta}_{i}\in\mathbb{C}^{2r\times2r}$ and $\bm{\varOmega}_{i}\in\mathbb{C}^{2r\times2(N-r)}$,
respectively. Accordingly, the third term in Eq. (\ref{eq: concentrated augmented likelihood perm})
is equivalently written as $-\frac{1}{2}\,\bm{m}_{c,i}^{\dagger}\,\bm{R}_{c,i}^{-1}\,\bm{m}_{c,i}$,
where $\bm{m}_{c,i}\triangleq(\bm{m}_{i})_{2r+1:2N}$. Furthermore,
it is also apparent that $\bm{R}_{c,i}$ is in the form $\bm{R}_{c,i}=(\bm{\Sigma}_{c,i}+\sigma_{J}^{2}\,\bm{I}_{2(N-r)})$,
where $\bm{\Sigma}_{c,i}\triangleq(\bm{\Gamma}^{\dagger}\bm{\Sigma}_{A,i}\bm{\Gamma})_{(2r+1:2N)}$.
Therefore $\bm{R}_{c,i}^{-1}$ has the eigenvalue decomposition $\bm{R}_{c,i}^{-1}=\bm{U}_{c,i}\,[\bm{\Lambda}_{c,i}+\sigma_{J}^{2}\,\bm{I}_{2(N-r)}]^{-1}\,\bm{U}_{c,i}^{\dagger}$.
Consequently, Eq.~(\ref{eq: concentrated augmented likelihood})
can be expressed as
\begin{align}
\ln[p_{\mathrm{igmm}}(\breve{\bm{s}}_{i};\hat{\underline{\bm{\zeta}}^{'}},\sigma_{J}^{2})]= & -N\,\ln\pi-\frac{1}{2}\sum_{n=1}^{2N}\ln[\lambda_{A,i,n}+\sigma_{J}^{2}]\nonumber \\
 & -\frac{1}{2}\sum_{\ell=1}^{2(N-r)}\frac{|v_{i,\ell}|^{2}}{\lambda_{c,i,\ell}+\sigma_{J}^{2}}\,,\label{eq: likelihood_sj2}
\end{align}
 where $\lambda_{A,i,n}$ and $\lambda_{c,i,\ell}$ are the eigenvalues
of $\bm{\Sigma}_{A,i}$ and $\bm{\Sigma}_{c,i}$, respectively. Also,
in Eq. (\ref{eq: likelihood_sj2}) and we have denoted with $v_{i,\ell}$
the $\ell$th element of $\bm{v}_{i}\triangleq(\bm{U}_{c,i}^{\dagger}\,\bm{m}_{c,i})$.
We also remark that, because of $\bm{\Sigma}_{A,i}$ definition, the
eigenvalues $\lambda_{A,i,n}$ \emph{are equal to those of }$\bm{\Sigma}_{\underline{\bm{y}}|\mathcal{H}_{i}}$.
Eq. (\ref{eq: likelihood_sj2}) can now be easily differentiated w.r.t.
$\sigma_{J}^{2}$ and set to zero in order to find the stationary
points. This is achieved via the solution of the polynomial equation:
\begin{equation}
\sum_{n=1}^{2N}\frac{1}{\lambda_{A,i,n}+\sigma_{J}^{2}}=\sum_{\ell=1}^{2(N-r)}\frac{|v_{i,\ell}|^{2}}{(\lambda_{c,i,\ell}+\sigma_{J}^{2})^{2}}\label{eq: poly_eq}
\end{equation}
Clearly, given a set of stationary points (to which we must add the
boundary solution $\hat{\sigma}_{J,i}^{2}=0$) say it $\hat{\sigma}_{J,i}^{2}(s)$,
the argument corresponding to the maximum likelihood of Eq. (\ref{eq: likelihood_sj2})
is chosen as the actual $\hat{\sigma}_{J,i}^{2}$, that is $\hat{\sigma}_{J,i}^{2}\triangleq\arg\max_{\hat{\sigma}_{J}^{2}(s)\geq0}\ln[p(\breve{\bm{s}}_{i};\hat{\underline{\bm{\zeta}}^{'}},\hat{\sigma}_{J}^{2}(s))]$.
This is also implied by the objective function $\ln[p_{\mathrm{igmm}}(\breve{\bm{s}}_{i};\hat{\underline{\bm{\zeta}}^{'}},\sigma_{J}^{2})]\rightarrow-\infty$
as $\sigma_{J}^{2}$ tends to $+\infty$. Finally, IGMM-GLR statistic
is evaluated as
\begin{gather}
\Lambda_{{\scriptscriptstyle \mathrm{IGMM-GLR}}}\triangleq-\frac{1}{2}\sum_{n=1}^{2N}\ln\left[\frac{\lambda_{A,1,n}+\hat{\sigma}_{J,1}^{2}}{\lambda_{A,0,n}+\hat{\sigma}_{J,0}^{2}}\right]\nonumber \\
-\frac{1}{2}\left\{ \sum_{n=1}^{2(N-r)}\frac{|v_{1,\ell}|^{2}}{\lambda_{c,1,\ell}+\hat{\sigma}_{J,1}^{2}}-\sum_{n=1}^{2(N-r)}\frac{|v_{0,\ell}|^{2}}{\lambda_{c,0,\ell}+\hat{\sigma}_{J,0}^{2}}\right\} \,.\label{eq: IGMM-GLRT statistic}
\end{gather}
The procedure for evaluation of IGMM-GLR statistic is summarized in
Alg. \ref{alg:IGMM-GLRT}. 

\begin{algorithm}
\texttt{Input: Evaluate $\bm{\Sigma}_{A,i}=(\bar{\bm{U}}_{J}^{\dagger}\,\bm{\Sigma}_{\underline{\bm{y}}|\mathcal{H}_{i}}\,\bar{\bm{U}}_{J})$,
$\bm{\Sigma}_{c,i}=(\bm{\Gamma}\,\bm{\Sigma}_{A,i}\,\bm{\Gamma}^{\dagger})_{(2r+1:2N)}$
and $\bm{\mu}_{i}=\widetilde{\bm{A}}(\bm{\theta})\,\bm{\rho}_{i}$.}

\texttt{(a) Given the vector $\bm{y}_{s}$, for each hypothesis $\mathcal{H}_{i}$:}
\begin{enumerate}
\item \texttt{Compute $\breve{\bm{s}}_{i}=\bm{U}_{J}^{\dagger}\,(\bm{y}_{s}-\bm{\mu}_{i})$;}
\item \texttt{Build the augmented vector $\underline{\breve{\bm{s}}_{i}}$
and evaluate $\bm{m}_{i}=\bm{\Gamma}\,\underline{\breve{\bm{s}}_{i}}$;}
\item \texttt{Obtain $\bm{m}_{c,i}=(\bm{m}_{i})_{2r+1:2N}$ and evaluate
$\bm{v}_{i}=\bm{U}_{c,i}^{\dagger}\,\bm{m}_{c,i}$;}
\item \texttt{Solve the polynomial equation in Eq.~(\ref{eq: poly_eq})
and take only solutions $\in\mathbb{R}^{+}$ plus $\hat{\sigma}_{J,i}^{2}=0$,
say it $\hat{\sigma}_{J,i}^{2}[s]$;}
\item \texttt{Obtain $\hat{\sigma}_{J,i}^{2}$ as}~\\
\texttt{ $\hat{\sigma}_{J,i}^{2}=\arg\max_{\sigma_{J,i}^{2}[s]}\ln[p(\breve{\bm{s}}_{i};\hat{\underline{\bm{\zeta}}^{'}},\sigma_{J,i}^{2}[s])]$;}
\end{enumerate}
\texttt{(b) Evaluate $\Lambda_{{\scriptscriptstyle \mathrm{IGMM-GLR}}}$
in Eq.~(\ref{eq: IGMM-GLRT statistic}).}

\caption{IGMM-GLR statistic evaluation. \label{alg:IGMM-GLRT}}
\end{algorithm}

\subsection{Asymptotic equivalences in the presence of jammer\label{subsec: Asymptotic Equivalences Jammer}}

Hereinafter, we will turn our attention to asymptotic equivalence
properties of fusion rules which deal with the case of jammer presence,
specularly as in Sec. \ref{subsec: Asymptotic Eq no sub}.

We first observe that, in the presence of jammer interference, it
is not difficult to show that a similar statement as that in Lem.
\ref{lem: Exact NLOS} does not hold, since there is a different design
criterion between NLOS-GLRT and IS/IGMM-GLRT. Indeed, the former is
obtained by exploiting a monotonic concentrated LLR (under NLOS assumption,
after Gaussian moment matching and implicit estimation of $\bm{\zeta}$);
these assumptions allow avoiding the estimation of $\sigma_{J}^{2}$.
Therefore, NLOS-GLRT cannot be interpreted as a GLRT-like procedure
in a strict sense, since it implicitly estimates only $\bm{\zeta}$.
On the other hand, IGMM-GLRT and IS-GLRT rules are both constructed
on an estimate $\hat{\sigma}_{J}^{2}$. Therefore, w\emph{e cannot
expect the three rules to have identical performance in a NLOS setting},
as opposed to the ``interference-free'' scenario. However, an intuitive
argument on their NLOS behaviour can be drawn by analyzing the forms
of IS-GLR (cf. Eq. (\ref{eq: IS-GLRT})) and IGMM-GLR (cf. Eq. (\ref{eq: IGMM-GLRT statistic}))
under the aforementioned assumption. Indeed, by assuming that the
Rician factors $\kappa_{k}\rightarrow0$, produces (after lengthy
manipulations):
\begin{gather}
\Lambda=\begin{cases}
N\,\ln\frac{\sigma_{a}^{2}}{\sigma_{b}^{2}}-\frac{\left\Vert \bm{r}_{0}\right\Vert ^{2}}{\sigma_{b}^{2}}+\frac{\left\Vert \bm{r}_{0}\right\Vert ^{2}}{\sigma_{a}^{2}} & \:\mathrm{if}\,\frac{\left\Vert \bm{r}_{0}\right\Vert ^{2}}{N}<\sigma_{a}^{2}\\
N\,\ln\frac{\left\Vert \bm{r}_{0}\right\Vert ^{2}}{N\sigma_{b}^{2}}-\frac{\left\Vert \bm{r}_{0}\right\Vert ^{2}}{\sigma_{b}^{2}}+N & \:\mathrm{if}\,\sigma_{a}^{2}\leq\frac{\left\Vert \bm{r}_{0}\right\Vert ^{2}}{N}<\sigma_{b}^{2}\\
0 & \:\mathrm{if}\,\frac{\left\Vert \bm{r}_{0}\right\Vert ^{2}}{N}\geq\sigma_{b}^{2}
\end{cases}\label{eq: NLOS - IS-IGMM GLRT}
\end{gather}
where $\sigma_{a}^{2}<\sigma_{b}^{2}$ and their expressions are $\sigma_{a}^{2}=\sigma_{n,0}^{2}=\sum_{k=1}^{K}\beta_{k}P_{F,k}+\sigma_{w}^{2}$
(resp. $\sigma_{b}^{2}=\sigma_{n,1}^{2}=\sum_{k=1}^{K}\beta_{k}P_{D,k}+\sigma_{w}^{2}$)
for IGMM-GLR and $\sigma_{a}^{2}=\sigma_{w}^{2}$ (resp. $\sigma_{b}^{2}=\bar{\sigma}_{e}^{2}(\bm{1}_{K})=\sum_{k=1}^{K}\beta_{k}+\sigma_{w}^{2}$)
for IS-GLR, respectively. By looking at Eq. (\ref{eq: NLOS - IS-IGMM GLRT}),
it is apparent that both the statistics are increasing functions of
$\left\Vert \bm{r}_{0}\right\Vert ^{2}$ (i.e., the energy of the
received signal $\bm{y}_{s}$ after projecting out the LOS part of
the jammer interference) within $[0,\sigma_{b}^{2}]$. Therefore,
the higher $\sigma_{b}^{2}$ the more the statistic function in Eq.~(\ref{eq: NLOS - IS-IGMM GLRT})
will be safely approximated by an increasing function of $\left\Vert \bm{r}_{0}\right\Vert ^{2}$.
Additionally, every statistic being an increasing function of $\left\Vert \bm{r}_{0}\right\Vert ^{2}$
will experience the same performance as the NLOS-GLRT (we recall that
such test is constructed simply comparing $\left\Vert \bm{r}_{0}\right\Vert ^{2}$
to a suitable threshold, cf. Eq. (\ref{eq: NLOS-GLRT})). Such test
is obtained without explicitly estimating $\sigma_{J}^{2}$ and by
claiming uniformly most powerfulness after moment matching of the
statistic $\left\Vert \bm{r}_{0}\right\Vert ^{2}$. The use of this
test allows avoiding a performance loss attributed to the fact that,
under a NLOS assumption, we are testing (after moment matching)
\begin{gather}
\begin{cases}
(\sigma_{n,0}^{2}+\sigma_{J}^{2}) & \,\;\mathrm{under}\,\mathcal{H}_{0}\\
(\sigma_{n,1}^{2}+\sigma_{J}^{2}) & \,\;\mathrm{under}\,\mathcal{H}_{1}
\end{cases}
\end{gather}
with $\sigma_{J}^{2}$ being unknown. Clearly, if we are faced to
estimate $\sigma_{J}^{2}$ under the condition $\sigma_{J}^{2}\geq(\sigma_{n,1}^{2}-\sigma_{n,0}^{2})$,
discrimination among the two hypotheses is not achievable. Indeed,
the uncertainty interval of $\sigma_{J}^{2}$ (i.e., $[0,+\infty)$)
produces overlapping intervals for the overall variance under both
hypotheses (i.e., $[\sigma_{n,0}^{2},+\infty)$ and $[\sigma_{n,1}^{2},+\infty)$,
respectively) and therefore, when the aforementioned condition is
satisfied, the correct hypothesis cannot be declared on the basis
of a simple variance estimation. Additionally, since $\sigma_{b}^{2}$
is higher for IS-GLR than for IGMM-GLR (as $P_{D,k}\leq1$, $k\in\mathcal{K}$),
we can expect IS-GLRT to perform \emph{better} than IGMM-GLRT in a
NLOS WSN situation, especially when $\sigma_{J}^{2}$ becomes large
(which is either the case of a jammer emitting a high power signal
or experiencing mostly a NLOS channel condition).

Finally, we show that an analogous form of Lem. \ref{lem: IS assumption}
holds for IS-GLRT and IGMM-GLRT in a setup with an operating jammer,
as stated hereinafter.
\begin{lem}
Under ``IS'' assumption, IGMM-GLRT rule is statistically equivalent
to IS-GLRT rule (and thus attains exact GLRT performance).
\end{lem}
\begin{IEEEproof}
Clearly, under IS assumption, IS-GLRT is statistically equivalent
to the exact GLRT in Eq. (\ref{eq: GLRT exploited}), by construction.
Then, we need only to show that IGMM-GLRT is statistically equivalent
to IS-GLRT. Indeed, under IS assumption, $\mathbb{E}\{\bm{x}|\mathcal{H}_{1}\}=\bm{1}_{K}$,
$\mathbb{E}\{\bm{x}|\mathcal{H}_{0}\}=\bm{0}_{K}$ and $\bm{\Sigma}_{\bm{x}|\mathcal{H}_{i}}=\bm{O}_{K}$
hold, respectively. Therefore, the second order characterization needed
for IGMM-GLRT in Eqs. (\ref{eq: IGMM-GLRT char 1}) and (\ref{eq: IGMM-GLRT char 2})
reduces to:
\begin{gather}
\mathbb{E}\{\bm{y}_{s}|\mathcal{H}_{i}\}=\bm{\mu}_{i}+\bm{A}_{J}(\bm{\phi})\,\bm{\zeta}\nonumber \\
\bm{\Sigma}_{\bm{y}_{s}|\mathcal{H}_{i}}=(\sigma_{e,i}^{2}+\sigma_{J}^{2})\,\bm{I}_{N}\qquad\bar{\bm{\Sigma}}_{\bm{y}_{s}|\mathcal{H}_{i}}=\bm{O}_{N}\label{eq: Char IGMM-GLRT - IS assumption}
\end{gather}
where the equalities $\sigma_{e,1}^{2}=\sigma_{e}^{2}(\bm{1}_{K})$,
$\sigma_{e,0}^{2}=\sigma_{w}^{2}$, $\bm{\mu}_{1}=\widetilde{\bm{A}}(\bm{\theta})\,\bm{1}_{K}$
and $\bm{\mu}_{0}=\bm{0}_{N}$ hold, respectively. It is apparent
that the simplified characterization in Eqs. (\ref{eq: Char IGMM-GLRT - IS assumption})
coincides with that in Eq. (\ref{eq: Char- IS GLRT}). Since both
rules are obtained with a GLRT-like approach, this proves their statistical
equivalence.
\end{IEEEproof}
Then, when sensors are ideal, IGMM-GLRT rule will be statistically
equivalent to IS-GLRT (viz. GLRT) rule, as there is no covariance
structure change between the two hypotheses. On the other hand, we
expect that when the WSN operates with ``good-quality'' sensors,
NLOS-GLRT will experience some performance loss with respect to IS-GLRT
and IGMM-GLRT rules, since it does not exploit the LOS part of the
sensors channel vectors.

\section{Complexity analysis\label{sec: Complexity analysis}}

In Tab. \ref{tab: Computational complexity} we compare the computational
complexity of the proposed rules, where $\mathcal{O}(\cdot)$ indicates
the usual Landau notation (i.e., the order of complexity). The results
underline the computations required whenever each new $\bm{y}$ is
transmitted (assuming static parameters pre-computed and stored in
a suitable memory). First, as previously remarked, it is apparent
that the optimum rule (i.e. the LLR) is unfeasible, especially when
$K$ is very large. Differently, all the proposed rules have polynomial
complexity w.r.t both $K$ and $N$ (as well as $r$, when jammer-robust
rules are considered). The computational complexity of IS rule is
mainly given by the computation of the scalar product and energy needed
to evaluate Eq. (\ref{eq: IS rule}), while the dominant term in the
case of IS-GLRT is represented by the evaluation of the energy of
$\bm{r}_{0}$ and $\bm{r}_{1}$, respectively (recall that the orthogonal
projector of interference can be written as $\bm{P}_{\bm{A}_{J}(\bm{\phi})}^{\perp}=\bm{U}_{J,\perp}\bm{U}_{J,\perp}^{\dagger}$,
where $\bm{U}_{J,\perp}$ collects the last $(N-r)$ columns of the
eigenvector matrix $\bm{U}_{J}$). Similar considerations (as IS rule)
hold for NLOS (which simply requires $\left\Vert \bm{y}\right\Vert ^{2}$),
whereas NLOS-GLRT similarly (as IS-GLRT) requires first a projection
operation, that is, evaluation of $\bm{P}_{\bm{A}_{J}(\bm{\phi})}^{\perp}\bm{y}_{s}$.
Furthermore, a linear dependence with $N$, as IS and NLOS rules,
holds for WL rules (see Eq. (\ref{eq: WL generical})). Differently,
IGMM rule is based on the computation of a quadratic form of $\bm{y}$,
which leads to $\mathcal{O}(N^{2})$ complexity. A higher complexity
is also required by IGMM-GLRT, whose dominant terms are given by:
($i$) the computation of $\bm{v}_{i}$ (see definition provided in
Sec. \ref{subsec:IGMM-GLRT-rule}) and the solution to a polynomial
equation of order $p_{\mathrm{ord}}\triangleq2N+4(N-r)-1$. The solution
is known to have a complexity $\mathcal{O}(p_{\mathrm{ord}}^{4}\tau^{2})$
(e.g. following Sturm approach \cite{Atallah2009}), where $\tau$
is a parameter related to the bit resolution of the maximum value
among the known coefficients.

\begin{table}
\caption{Computational complexity of the considered rules; $p_{\mathrm{ord}}\triangleq2N+4(N-r)-1$.\label{tab: Computational complexity}}

\centering{}%
\begin{tabular}{c||l}
\hline 
\noalign{\vskip\doublerulesep}
\textbf{Fusion Rule} & \textbf{Complexity for each realization of $\bm{y}$ }\tabularnewline[\doublerulesep]
\hline 
\noalign{\vskip\doublerulesep}
\hline 
\noalign{\vskip\doublerulesep}
Optimum (LLR) & $\mathcal{O}(N\,2^{K})$\tabularnewline[\doublerulesep]
\hline 
\noalign{\vskip\doublerulesep}
IS {[}IS-GLRT{]}  & $\mathcal{O}(N)$ {[} $\mathcal{O}(N(N-r))$ {]}\tabularnewline[\doublerulesep]
\hline 
\noalign{\vskip\doublerulesep}
NLOS {[}NLOS-GLRT{]}  & $\mathcal{O}(N)$ {[} $\mathcal{O}(N(N-r))$ {]}\tabularnewline[\doublerulesep]
\hline 
\noalign{\vskip\doublerulesep}
WL  & $\mathcal{O}(N)$\tabularnewline[\doublerulesep]
\hline 
\noalign{\vskip\doublerulesep}
IGMM {[}IGMM-GLRT{]}  & $\mathcal{O}(N^{2})$ {[}$\mathcal{O}(N\,(N-r)+p_{\mathrm{ord}}^{4}\,\tau^{2})${]}\tabularnewline[\doublerulesep]
\hline 
\end{tabular}
\end{table}

\section{Simulation results\label{sec: Simulation results}}

\subsection{Setup description and measures of performance}

We consider sensors deployed in a 2-D circular area around the DFC
(placed in the origin, whose cartesian coordinates are denoted as
($x_{\mathrm{dfc}},y_{\mathrm{dfc}})$) with radius $r_{\mathrm{max}}=1000\,\unit{m}$.
Sensors are located uniformly at random (in Cartesian coordinates,
denoted as $(p_{x,k},p_{y,k})$, $k\in\mathcal{K}$) and we assume
that no sensor is closer to the DFC than $r_{\mathrm{min}}=100\,\unit{m}$.
The large-scale fading is modelled via $\beta_{k}=\xi{}_{k}(\frac{r_{\mathrm{min}}}{r_{k}})^{L}$,
where $\xi_{k}$ is a log-normal random variable, i.e., $10\log_{10}(\xi_{k})\sim\mathcal{N}(\mu_{\mathrm{P}},\sigma_{\mathrm{P}}^{2})$,
where $\mu_{\mathrm{P}}$ and $\sigma_{\mathrm{P}}$ are the mean
and standard deviation in $\mathrm{dBm}$, respectively. Moreover,
$r_{k}$ denotes the distance between the $k$th sensor and the DFC
and $L$ represents the path-loss exponent (for our simulations, we
choose $L=2$). In the following, we assume $(\mu_{\mathrm{P}},\sigma_{\mathrm{P}})=(15,2)$
for the WSN. Additionally, we suppose that the DFC is equipped with
a half-wavelength spaced uniformly linear array and that $k$th sensor
is seen at the DFC as a point-like source, that is 
\begin{equation}
\bm{a}(\theta_{k})=\begin{bmatrix}1 & e^{j\pi\cos(\theta_{k})} & \cdots & e^{j\pi(N-1)\cos(\theta_{k})}\end{bmatrix}^{T}
\end{equation}
where clearly $\theta_{k}=\arccos[\frac{x_{\mathrm{dfc}}-p_{x,k}}{y_{\mathrm{dfc}}-p_{y,k}}]$.
A similar procedure is employed for the generation of jammer parameters,
with reference to a case of a jamming device distributed in angular
space. The sole difference is in the choice $(\mu_{\mathrm{P}},\sigma_{\mathrm{P}})=(25,2)$,
reflecting a non-negligible jammer power received by the DFC.

Also, the Rician factors of the sensors $\kappa_{k}$, $k\in\mathcal{K}$,
are uniformly generated within $[\mathrm{\kappa}_{\mathrm{min}},\mathrm{\kappa}_{\mathrm{max}}]$.
Such interval will be varied in order to generate three typical scenarios
corresponding to a WSN with ``LOS'', ``Intermediate'' and ``NLOS''
channel situations, in order to comprehensively test the proposed
fusion rules. More specifically, we will consider Rician factors generated
randomly as: ($i$) $[\mathrm{\kappa}_{\mathrm{min}},\mathrm{\kappa}_{\mathrm{max}}]=[10,20]\,(\mathrm{dB})$
(LOS scenario), ($ii$) $[\mathrm{\kappa}_{\mathrm{min}},\mathrm{\kappa}_{\mathrm{max}}]=[-10,10]\,(\mathrm{dB})$
(Intermediate scenario) and $[\mathrm{\kappa}_{\mathrm{min}},\mathrm{\kappa}_{\mathrm{max}}]=[-20,-10]\,(\mathrm{dB})$
(NLOS scenario). Similar reasoning is applied to the generation of
Rician factors for the jammer, where two different scenarios are also
considered: ($a$) $[\mathrm{\kappa}_{\mathrm{min}},\mathrm{\kappa}_{\mathrm{max}}]=[10,20]\,(\mathrm{dB})$
(LOS jammer) and ($b$) $[\mathrm{\kappa}_{\mathrm{min}},\mathrm{\kappa}_{\mathrm{max}}]=[-10,10]\,(\mathrm{dB})$
(weak-LOS jammer).

\begin{figure*}
\begin{centering}
\subfloat[LOS scenario for WSN. \label{fig: LOS scenario WSN}]{\includegraphics[width=0.75\paperwidth]{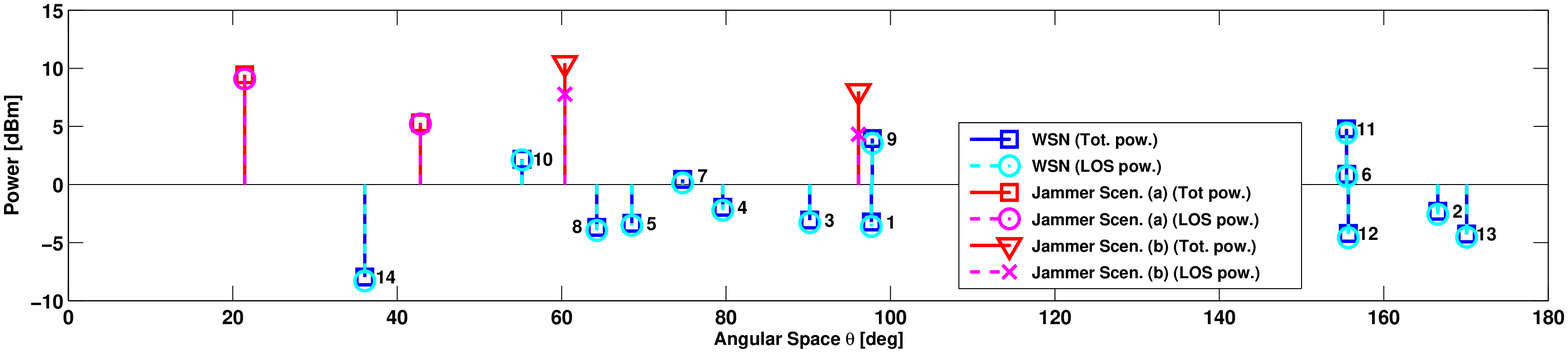}}
\par\end{centering}
\begin{centering}
\subfloat[Intermediate scenario for WSN.\label{fig: Intermediate scenario WSN}]{\includegraphics[width=0.75\paperwidth]{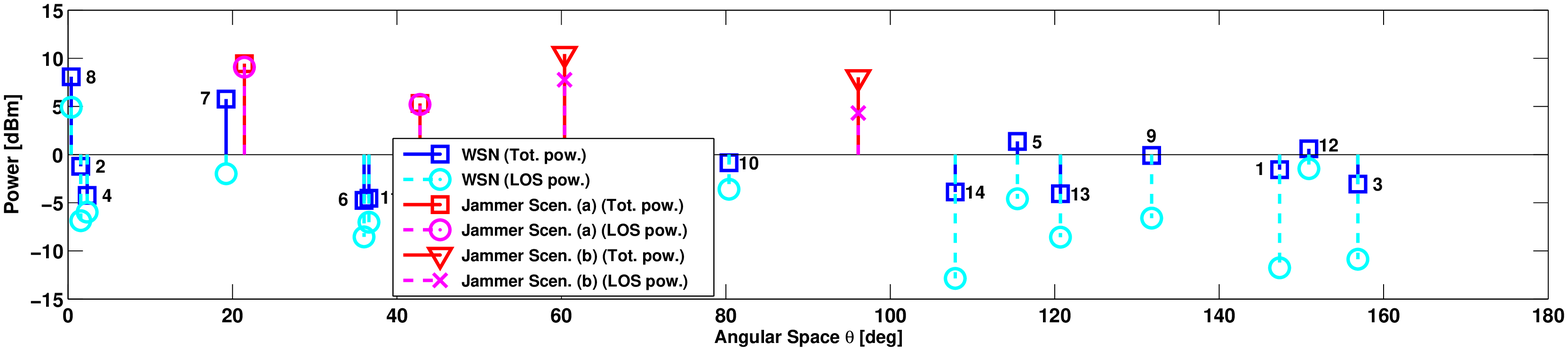}}
\par\end{centering}
\centering{}\subfloat[NLOS scenario for WSN.\label{fig: NLOS scenario WSN}]{\includegraphics[width=0.75\paperwidth]{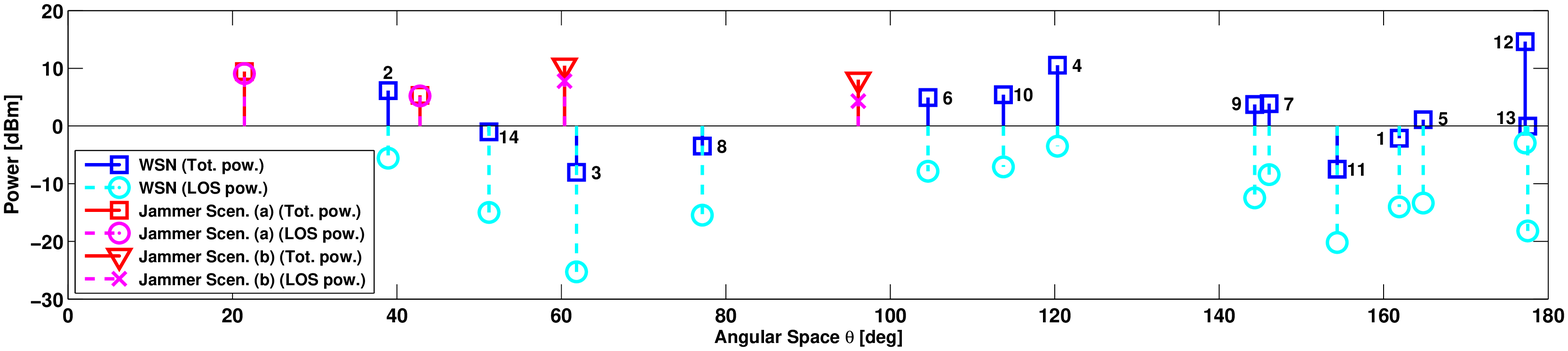}}\caption{Simulated setups for fusion rules comparison. Overall ($\beta_{k}$)
and LOS ($b_{k}^{2}\,\beta_{k}$) received power ($\mathrm{dBm}$)
per antenna at the DFC vs. $\theta$ ($\mathrm{deg}$) in a WSN with
$K=14$ sensors (blue ``$\square$'' and cyan ``$\circ$'' markers,
resp.). Each subfigure also reports the overall ($\beta_{\ell,J}$)
and LOS ($b_{\ell,J}^{2}\,\beta_{\ell,J}$) received power ($\mathrm{dBm}$)
per antenna at the DFC vs. $\theta$ ($\mathrm{deg}$) of a distributed
jammer with $r=2$ in scenarios ($a$) (LOS jammer scenario, red ``$\square$''
and magenta ``$\circ$'' markers, resp.) and ($b$) (weak-LOS jammer
scenario, red ``$\nabla$'' and magenta ``$\times$'' markers,
resp.). \label{fig: Scenarios for numerical comparison}}
\end{figure*}

The three generated WSN examples are shown in Fig.~\ref{fig: Scenarios for numerical comparison},
where the corresponding angles-of-arrival ($\theta_{k}$, $k\in\mathcal{K}$),
the averaged total received and LOS powers per antenna ($(\beta_{k},b_{k}^{2}\beta_{k})$,
$k\in\mathcal{K}$ ) are shown for the case of $K=14$ sensors. Also,
in each of the subfigures, we illustrate the corresponding DOAs ($\phi_{\ell}$,
$\mathcal{\ell=}1,\ldots,r$), the averaged total received and LOS
powers per antenna ($(\beta_{\ell,J},b_{\ell.J}^{2}\beta_{\ell,J})$,
$\mathcal{\ell=}1,\ldots,r$) of a jammer distributed in the angular
space with $r=2$, whose Rician factors are generated according to
scenarios ($a$) (LOS jammer scenario) and ($b$) (weak-LOS jammer
scenario), respectively. Finally, for simplicity we assume conditionally
i.i.d. decisions, that is $P(\bm{x}|\mathcal{H}_{i})=\prod_{k=1}^{K}P(x_{k}|\mathcal{H}_{i})$
with $(P_{1},P_{0})=(P_{D},P_{F})=(0.5,0.05)$. In this case, $\bm{\rho}_{i}=P_{i}\bm{1}_{K}$
and $\bm{\Sigma}_{\bm{x}|\mathcal{H}_{i}}=P_{i}(1-P_{i})\,\bm{I}_{K}$
hold, respectively.

The performances of the proposed rules are analyzed in terms of system
probabilities of false alarm and detection, defined respectively as
\begin{gather}
P_{F_{0}}\triangleq P(\Lambda>\gamma|H_{0}),\quad P_{D_{0}}\triangleq P(\Lambda>\gamma|H_{1}),
\end{gather}
with $\Lambda$ representing the statistic associated to the generic
fusion rule and $\gamma$ the corresponding threshold.

\subsection{Fusion Rules Comparison}

\begin{figure}
\includegraphics[width=0.9\columnwidth]{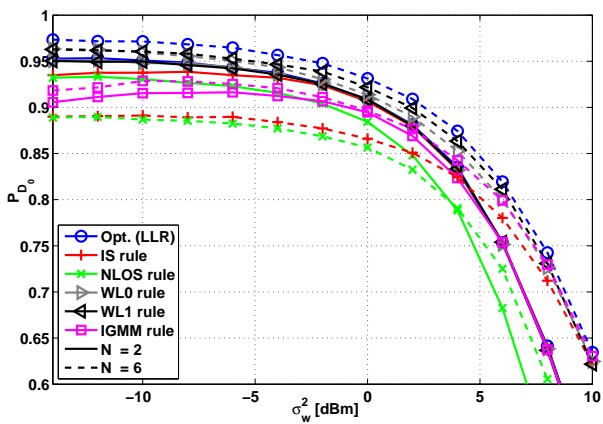}\caption{$P_{D_{0}}$ vs. $\sigma_{w}^{2}$ ($\mathrm{dBm}$) for a WSN with
$K=14$, LOS setup; $P_{F_{0}}=(0.01)$; $N\in\{2,6\}$ antennas at
the DFC.\label{fig: Pd0 vs sigma (LOS)}}
\end{figure}

\begin{figure}
\includegraphics[width=0.9\columnwidth]{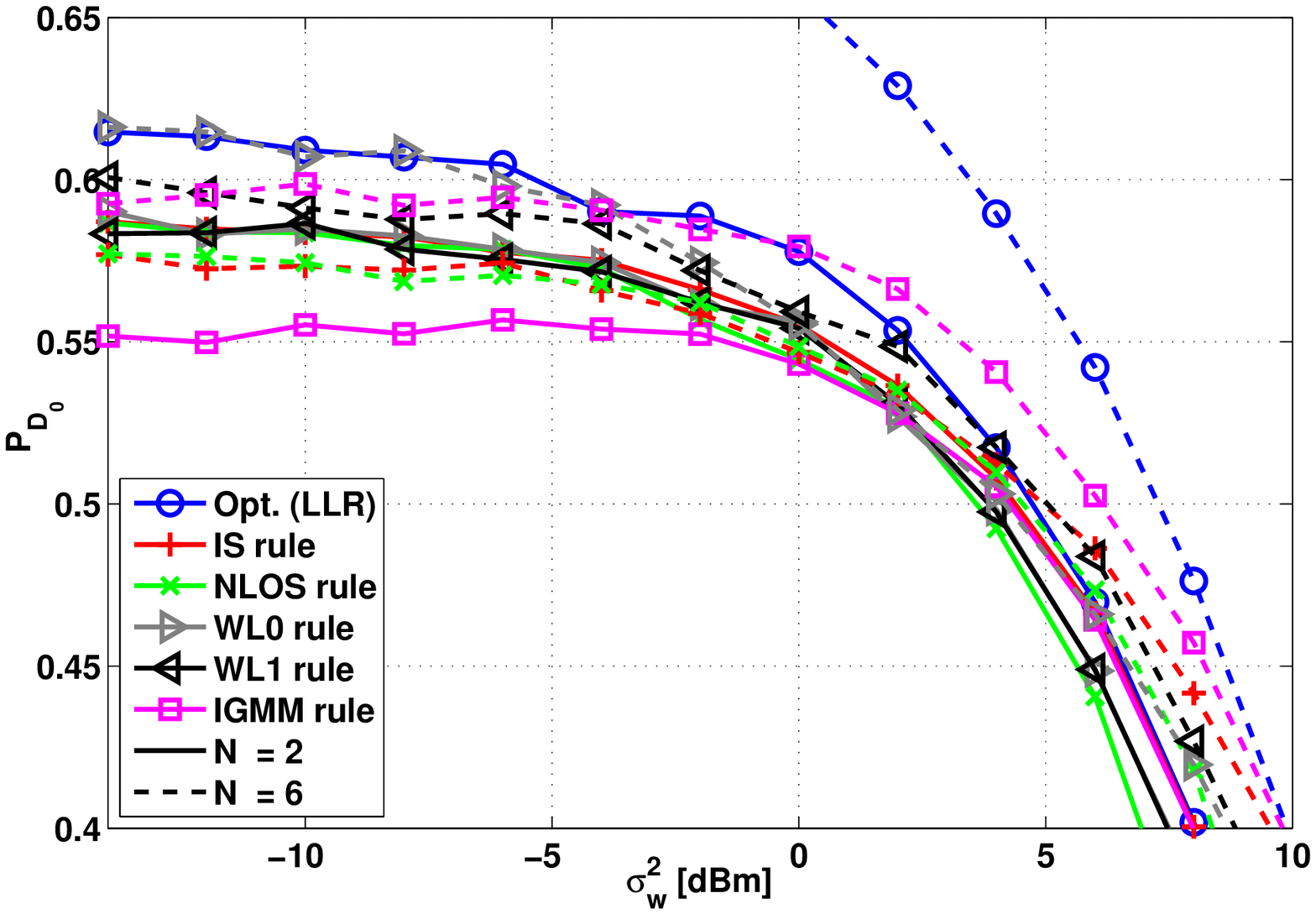}\caption{$P_{D_{0}}$ vs. $\sigma_{w}^{2}$ ($\mathrm{dBm}$) for a WSN with
$K=14$, Intermediate setup; $P_{F_{0}}=(0.01)$; $N\in\{2,6\}$ antennas
at the DFC.\label{fig: Pd0 vs sigma (Interm)}}
\end{figure}

\begin{figure}
\includegraphics[width=0.9\columnwidth]{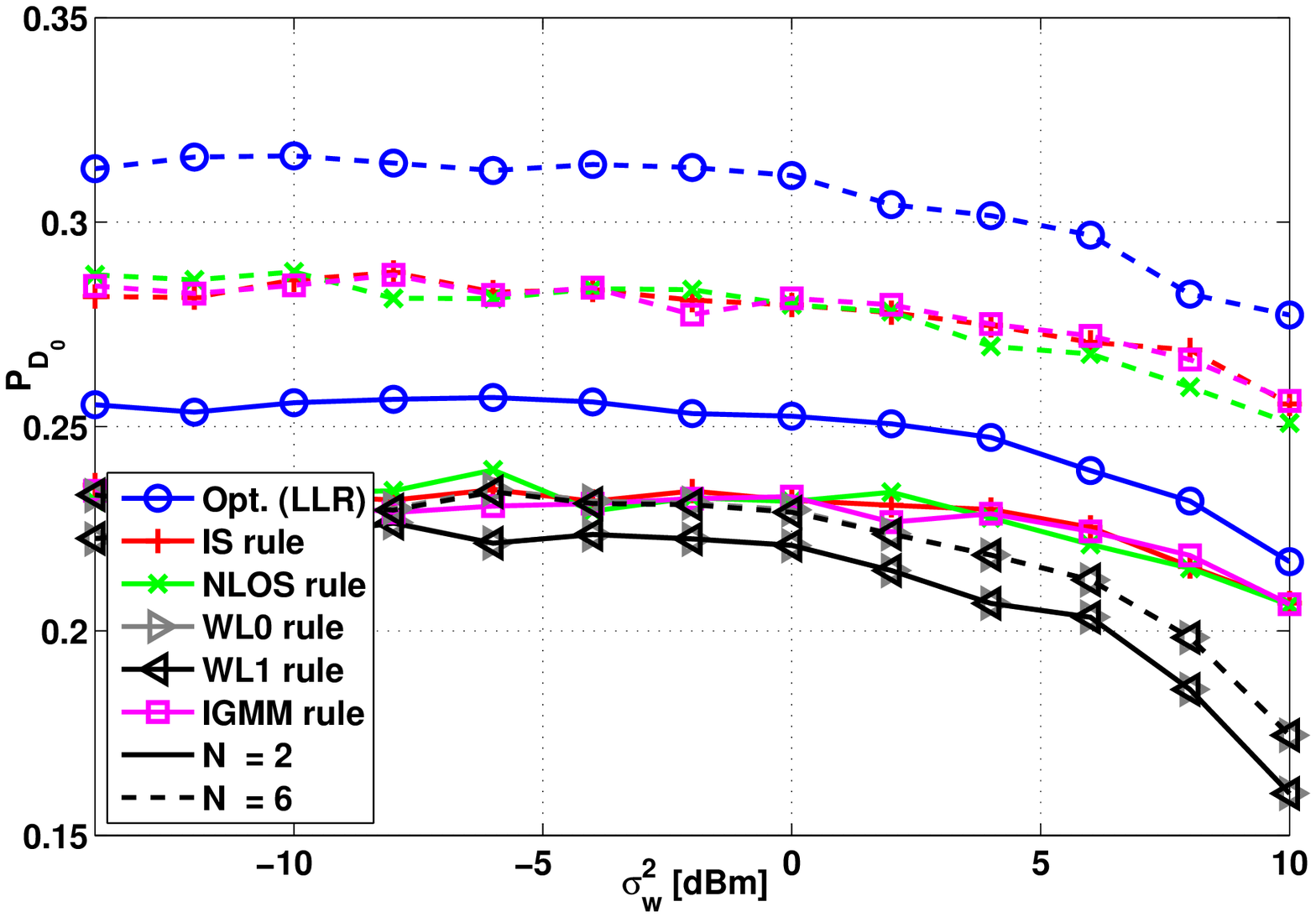}\caption{$P_{D_{0}}$ vs. $\sigma_{w}^{2}$ ($\mathrm{dBm}$) for a WSN with
$K=14$, NLOS setup; $P_{F_{0}}=(0.01)$; $N\in\{2,6\}$ antennas
at the DFC.\label{fig: Pd0 vs sigma (NLOS)}}
\end{figure}

\emph{$P_{D_{0}}$ vs. noise level $\sigma_{w}^{2}$ (No-interference):
}First, the scenario with \emph{no jammer} is addressed. In Figs.
\ref{fig: Pd0 vs sigma (LOS)}, \ref{fig: Pd0 vs sigma (Interm)}
and \ref{fig: Pd0 vs sigma (NLOS)}, we show $P_{D_{0}}$ vs. $\sigma_{w}^{2}$,
under the constraint $P_{F_{0}}=0.01$ for the ``LOS'', ``Intermediate''
and ``NLOS'' setups in Figs. \ref{fig: LOS scenario WSN}, \ref{fig: Intermediate scenario WSN}
and \ref{fig: NLOS scenario WSN}, respectively ($K=14$ sensors and
$N\in\{2,6\}$ antennas at the DFC). Clearly, LLR performs the best
among all the considered rules. Secondly, WL rules are very close
to the LLR in the ``LOS'' setup (indeed in the conditionally i.i.d.
case and at high SNR, for a LOS condition it approximately holds $\underline{\bm{z}}_{\,{\scriptscriptstyle \mathrm{WL}},i}\propto(\underline{\widetilde{\bm{A}}}(\bm{\theta})\,\underline{\widetilde{\bm{A}}}(\bm{\theta})^{\dagger})^{-1}\underline{\widetilde{\bm{A}}}(\bm{\theta})\bm{1}_{k}$,
that is WL rules both \emph{approximate} through right-pseudoinverse
operation a counting rule, being optimal in this specific scenario)
with increasing performance loss in the ``Intermediate'' and ``NLOS''
setups, respectively. Such a trend is in agreement with Lem. \ref{lem: Exact NLOS},
which states that as NLOS assumption is verified, the optimum statistic
should possess a dependence on $\left\Vert \bm{y}\right\Vert ^{2}$,
which is not the case of WL rules. Also, IGMM, IS and NLOS rules have
a performance behaviour in line with the asymptotic equivalences shown
in Sec. \ref{subsec: Asymptotic Eq no sub}. Clearly, NLOS setup is
such that performance of IGMM, IS and NLOS rules (almost) coincide.
On the other hand, in the LOS scenario, IS and IGMM rules are very
close (the ``weak-LOS'' assumption is almost satisfied), while NLOS
rule experiences a certain performance loss. Finally, we underline
that the benefit of improved number of antennas is only experienced
by LLR, WL and IGMM rules. Differently, NLOS and IS rules \emph{do}
benefit of a larger DFC array \emph{only} in the case of low SNR or
NLOS setup. This can be attributed to the fact that only in these
conditions there is no significant (pseudo-)covariance structure change
between the two hypotheses (see (\ref{eq: 2nd order char COV}) and
(\ref{eq: 2nd order char PCOV})). Then NLOS and IS rules, not exploiting
(at least) a second-order characterization of $\bm{y}|\mathcal{H}_{i}$,
are not able to benefit from increase of $N$ in the remaining cases.


\begin{figure}
\includegraphics[width=0.95\columnwidth]{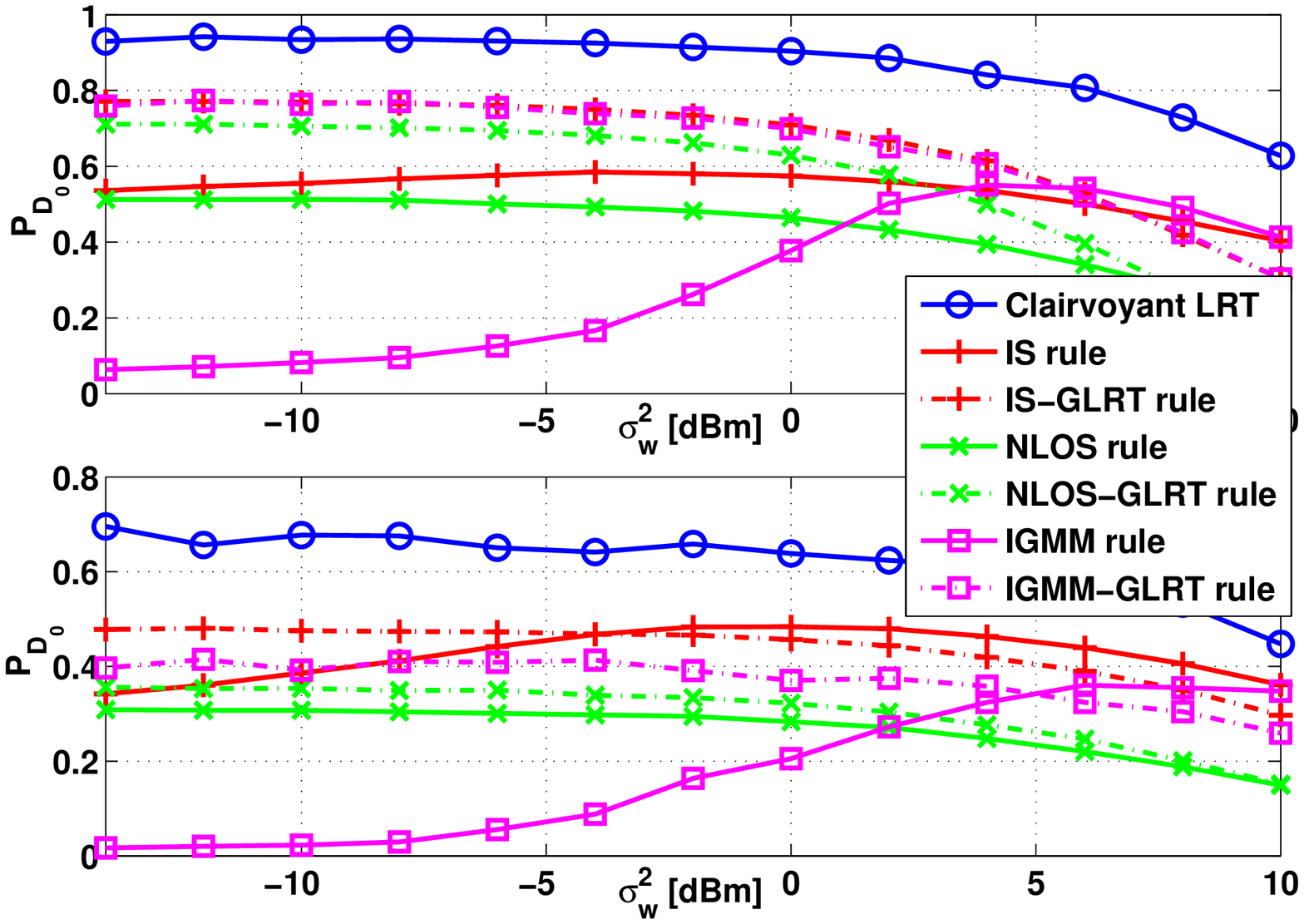}\caption{$P_{D_{0}}$ vs. $\sigma_{w}^{2}$ ($\mathrm{dBm}$) for a WSN with
$K=14$ sensors. LOS setup with jammer interference ($r=2$): top
figure - scenario ($a$) (LOS jam.); bottom figure - scenario ($b$)
(weak-LOS jam.); $P_{F_{0}}=0.01$. $N=6$ antennas at the DFC.\label{fig: Pd0 vs sigma (LOS-SUB)}}
\end{figure}

\begin{figure}
\includegraphics[width=0.95\columnwidth]{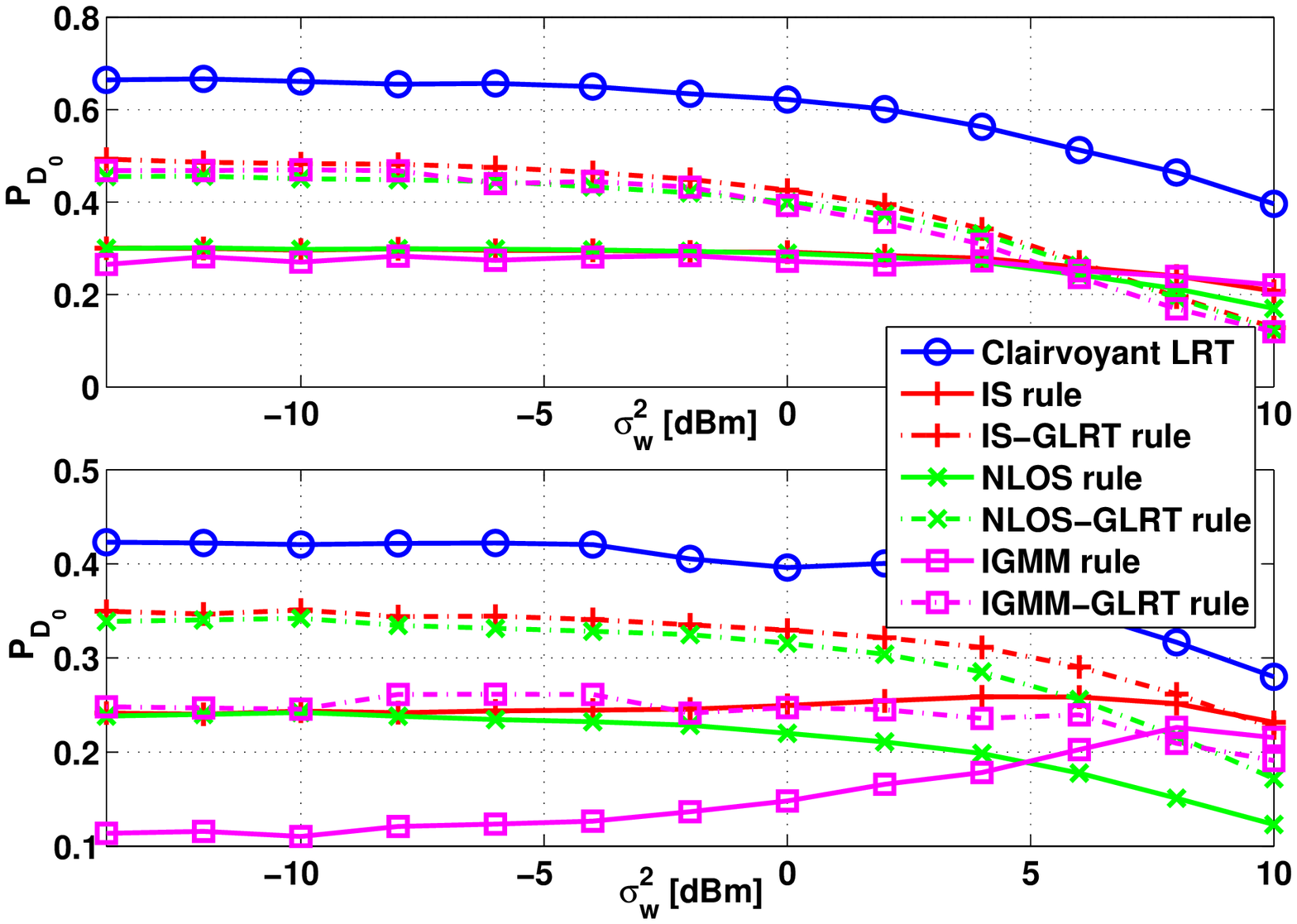}\caption{$P_{D_{0}}$ vs. $\sigma_{w}^{2}$ ($\mathrm{dBm}$) for a WSN with
$K=14$ sensors. Intermediate setup with jammer interference ($r=2$):
top figure - scenario ($a$) (LOS jam.); bottom figure - scenario
($b$) (weak-LOS jam.); $P_{F_{0}}=0.01$. $N=6$ antennas at the
DFC.\label{fig: Pd0 vs sigma (Interm-SUB)}}
\end{figure}

\begin{figure}
\includegraphics[width=0.95\columnwidth]{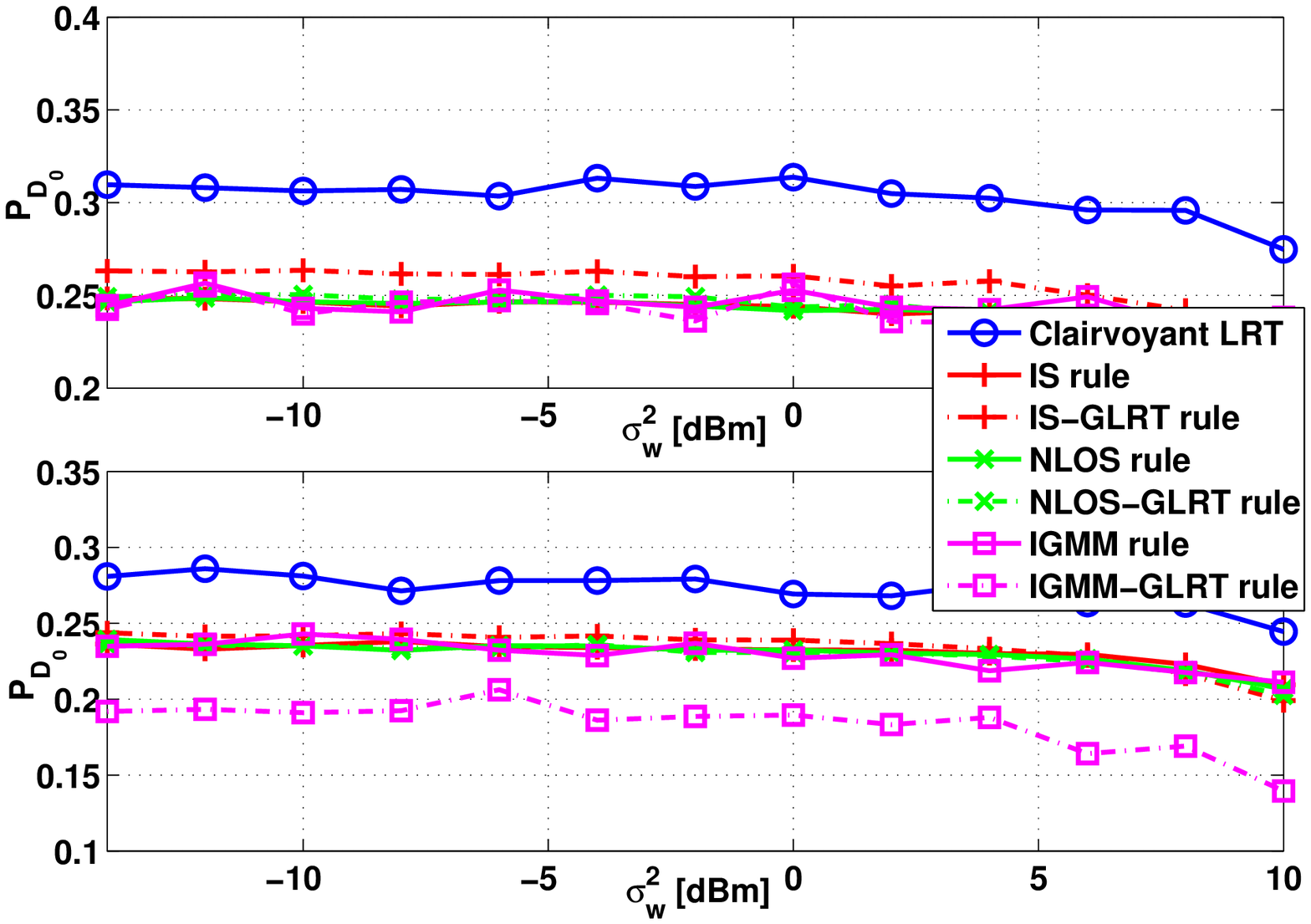}\caption{$P_{D_{0}}$ vs. $\sigma_{w}^{2}$ ($\mathrm{dBm}$) for a WSN with
$K=14$. NLOS setup with jammer interference ($r=2$): top figure
- scenario ($a$) (LOS jam.); bottom figure - scenario ($b$) (weak-LOS
jam.); $P_{F_{0}}=0.01$. $N=6$ antennas at the DFC.\label{fig: Pd0 vs sigma (NLOS-SUB)}}
\end{figure}

\emph{$P_{D_{0}}$ vs. noise level $\sigma_{w}^{2}$ (Interference):
}A similar scenario is shown in Figs. \ref{fig: Pd0 vs sigma (LOS-SUB)},
\ref{fig: Pd0 vs sigma (Interm-SUB)} and \ref{fig: Pd0 vs sigma (NLOS-SUB)},
where we show $P_{D_{0}}$ vs. $\sigma_{w}^{2}$, under the constraint
$P_{F_{0}}=0.01$ for the ``LOS'', ``Intermediate'' and ``NLOS''
setups in Figs. \ref{fig: LOS scenario WSN}, \ref{fig: Intermediate scenario WSN}
and \ref{fig: NLOS scenario WSN}, respectively ($K=14$, both jammer
scenarios considered), and $N=6$ antennas at the DFC. For the sake
of completeness, the performance of clairvoyant LRT are also reported
(cf. Eq. (\ref{eq: Clairvoyant LRT rule})). We first notice that
IS-GLRT, NLOS-GLRT and IGMM-GLRT outperform IS, NLOS and IGMM rules
(whose performance are obtained by ignoring the presence of the jamming
signal), respectively, unless there is a significant receive noise
$\sigma_{w}^{2}$ (i.e., low SNR); such trend is more apparent when
moving to a WSN-DFC channel which experiences a LOS scenario (cf.
Fig. \ref{fig: Pd0 vs sigma (LOS-SUB)}). Indeed, in such a case,
jammer interference suppression may come up at the expenses of (partial)
cancellation of some of the sensors contributions. Indeed, in a LOS
scenario and at low SNR, jammer interference suppression may not be
beneficial as the scenario is \emph{noise-dominated}. On the other
hand, in a LOS scenario and at high SNR, the problem becomes \emph{interference-dominated};
therefore an effective jammer signal suppression significantly improves
performance, even at the expenses of (partial) elimination of some
sensors contributions. The sole exception to these considerations
is represented by IGMM-GLRT in a NLOS WSN scenario (cf. Fig. \ref{fig: NLOS scenario WSN}),
where performance are observed to be worse than its interference-unaware
counterpart (i.e., IGMM rule) over all the $\sigma_{w}^{2}$ range
considered. Such evidence can be attributed to the overlapping of
unknown parameter support under the two hypotheses, due to $\sigma_{J}^{2}$
(cf. Sec. \ref{subsec: Asymptotic Equivalences Jammer}), which does
not allow to achieve satisfactory performance.


\begin{figure}
\includegraphics[width=0.95\columnwidth]{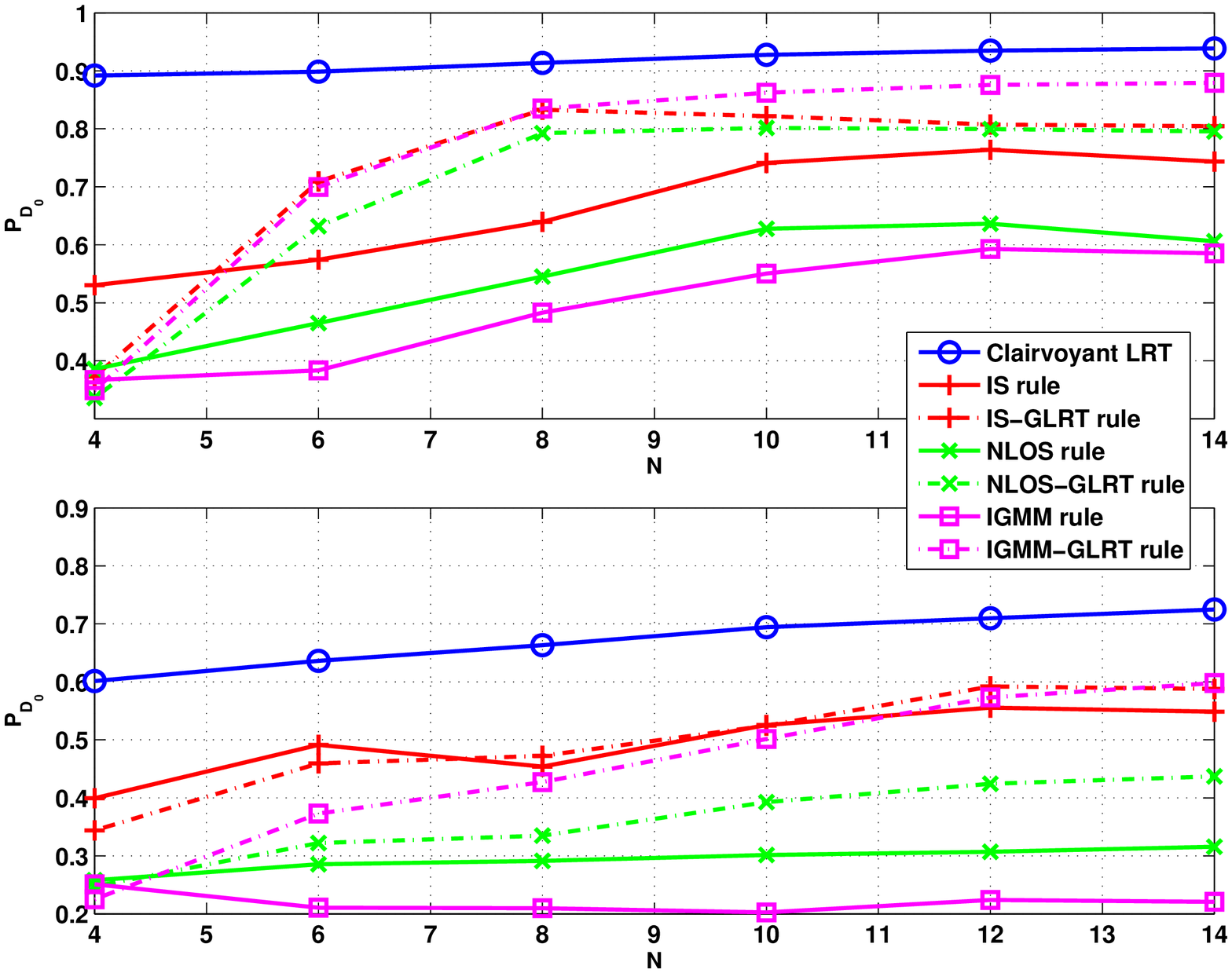}\caption{$P_{D_{0}}$ vs. $N$ for a WSN with $K=14$, $\sigma_{w}^{2}=0\,\mathrm{dBm}$.
LOS setup with jammer interference ($r=2$): top figure - scenario
($a$) (LOS jam.); bottom figure - scenario ($b$) (weak-LOS jam.);
$P_{F_{0}}=0.01$. \label{fig: Pd0 vs N (LOS-SUB)}}
\end{figure}

\begin{figure}
\includegraphics[width=0.95\columnwidth]{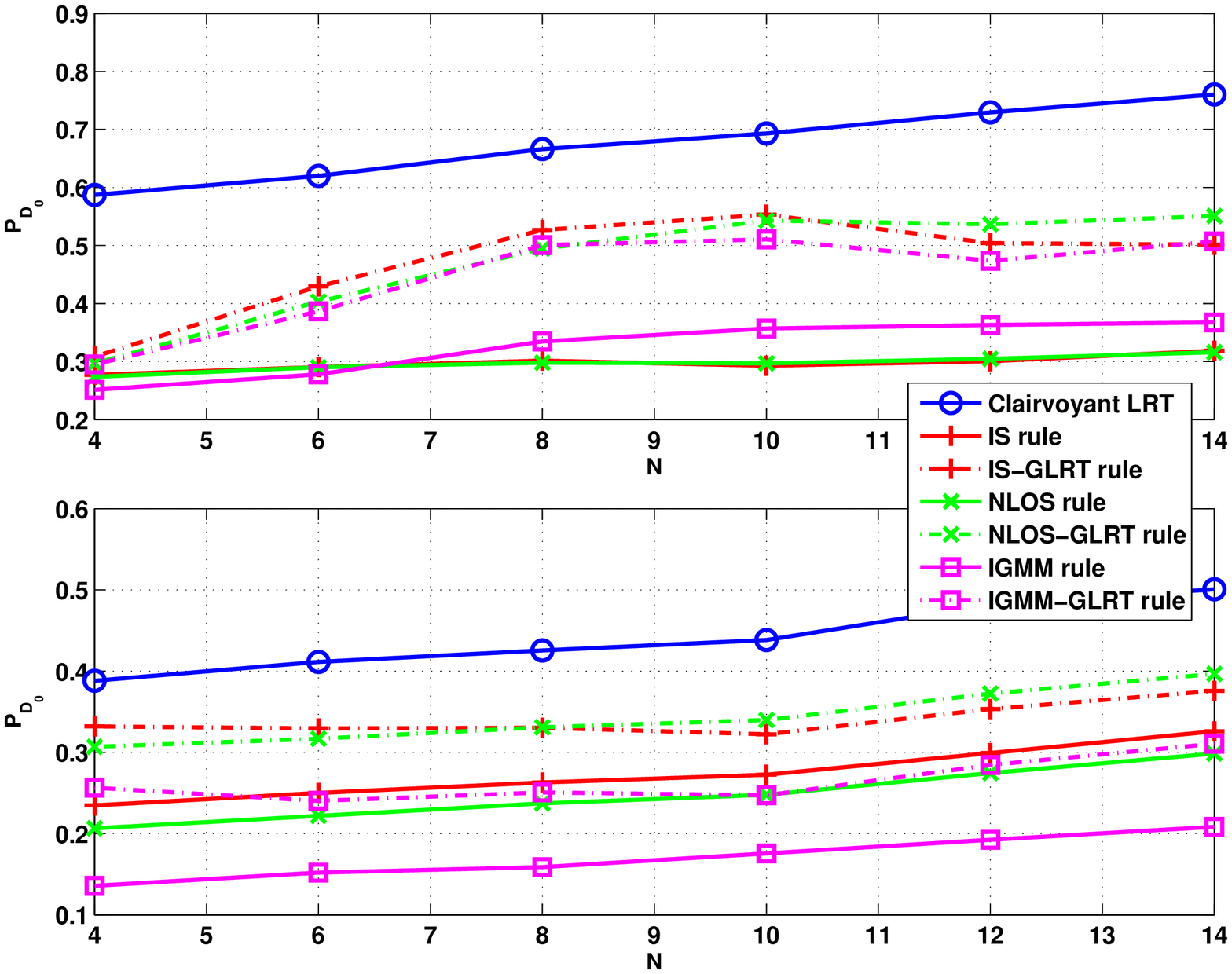}\caption{$P_{D_{0}}$ vs. $N$ for a WSN with $K=14$, $\sigma_{w}^{2}=0\,\mathrm{dBm}$.
Intermediate setup with jammer interference ($r=2$): top figure -
scenario ($a$) (LOS jam.); bottom figure - scenario ($b$) (weak-LOS
jam.); $P_{F_{0}}=0.01$.\label{fig: Pd0 vs N (Interm-SUB)}}
\end{figure}

\begin{figure}
\includegraphics[width=0.95\columnwidth]{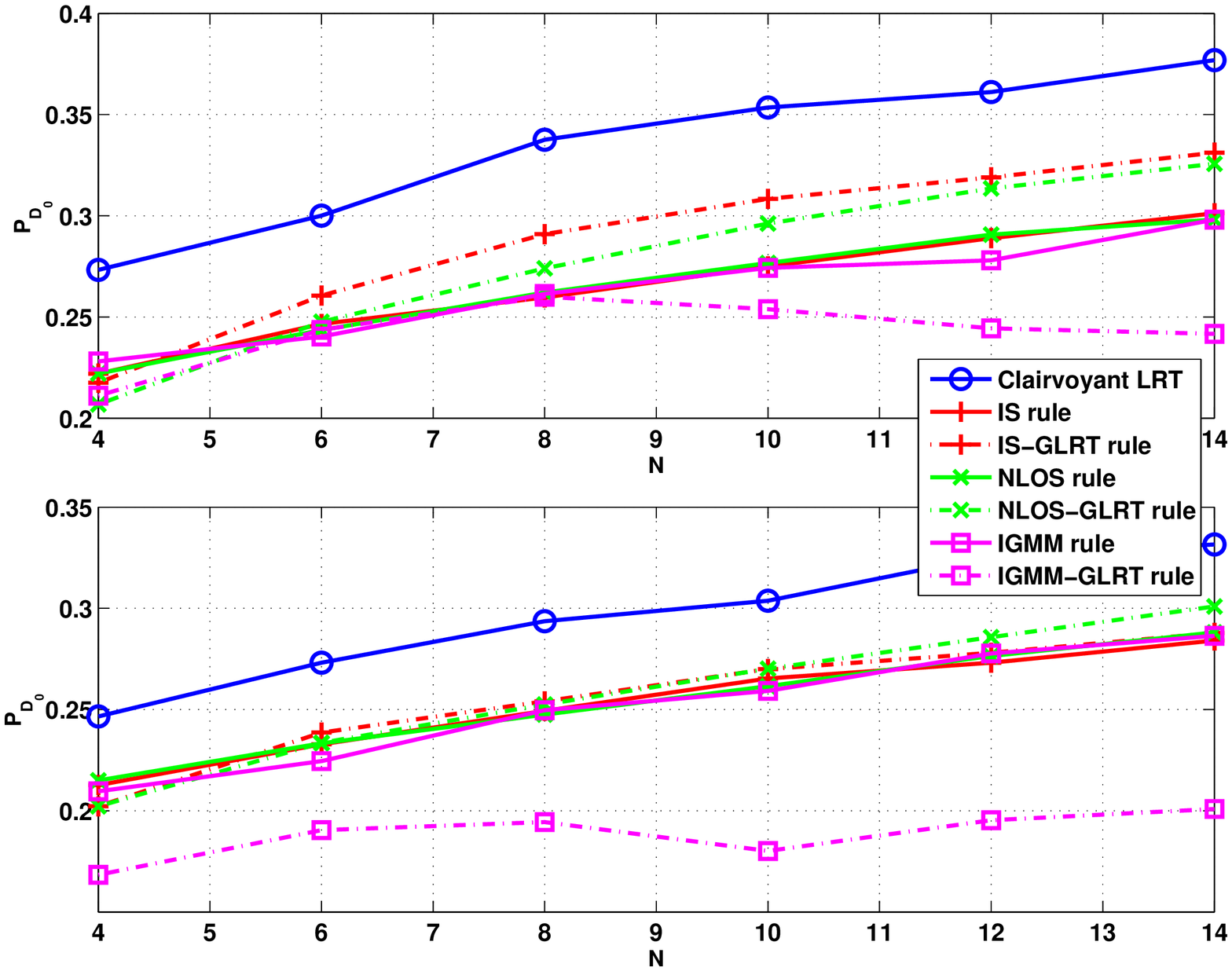}\caption{$P_{D_{0}}$ vs. $N$ for a WSN with $K=14$, $\sigma_{w}^{2}=0\,\mathrm{dBm}$.
NLOS setup with jammer interference ($r=2$): top figure - scenario
($a$) (LOS jam.); bottom figure - scenario ($b$) (weak-LOS jam.);
$P_{F_{0}}=0.01$.\label{fig: Pd0 vs N (NLOS-SUB)}}
\end{figure}

\emph{$P_{D_{0}}$ vs. number of antennas $N$ (Interference): }The
benefits of increasing the number of antennas on jammer suppression
capabilities for the designed rules are illustrated in Figs. \ref{fig: Pd0 vs N (LOS-SUB)},
\ref{fig: Pd0 vs N (Interm-SUB)}, and \ref{fig: Pd0 vs N (NLOS-SUB)},
respectively. More specifically, it is shown $P_{D_{0}}$ vs. $N$,
under the constraint $P_{F_{0}}=0.01$ and $\sigma_{w}^{2}=0\,\mathrm{dBm}$
for the ``LOS'', ``Intermediate'' and ``NLOS'' setups in Figs.
\ref{fig: LOS scenario WSN}, \ref{fig: Intermediate scenario WSN}
and \ref{fig: NLOS scenario WSN}, respectively. First of all, we
notice that $P_{D_{0}}$ for all the ``interference-aware'' fusion
rules increases with $N$. Furthermore, the gain with respect to their
corresponding ``interference-unaware'' counterparts improves as
well. This is true in the case of IS-GLRT and NLOS-GLRT for all the
scenarios considered, since the considered noise level $\sigma_{w}^{2}$
implies a moderate SNR and due to improved jamming-suppression capabilities
with higher $N$. Again, the only exception is given by IGMM-GLRT
in the NLOS case, given the overlapping of unknown parameter support
under both the hypotheses due to $\sigma_{J}^{2}$. By looking at
the specific example, in the LOS WSN scenario, the IGMM-GLRT has the
\emph{best trend} with the number of antennas, as significant pseudo-covariance
structure change in the hypotheses is implied in such scenario (therefore
second-order characterization of $\bm{y}|\mathcal{H}_{i}$ is beneficial).
Differently, in the ``Intermediate'' and ``NLOS'' WSN setups,
IS-GLRT and NLOS-GLRT represent the best alternatives, with IS-GLRT
slightly outperforming NLOS-GLRT.


\section{Conclusions\label{sec: Conclusions}}

In this paper we studied channel-aware DF in a WSN with interfering
sensors whose channels are modelled as Rician and their NLOS components
are not known at the DFC (i.e., they are not estimated), focusing
on anomaly detection problems. We developed five sub-optimal fusion
rules (i.e., IS, NLOS, WL and IGMM rules) in order to deal with the
exponential complexity of LRT. For the present setup, the following
performance trends have been observed:
\begin{itemize}
\item In a WSN with a LOS setup, WL rules represent the best (and more convenient)
alternative to the LLR, whereas the same rules suffer from severe
performance degradation in a NLOS setup. On the other hand, NLOS rule
is mainly appealing in a NLOS setup, as the IS rule, which also achieves
satisfactory performance in a weak-LOS condition. Indeed, in the latter
case they are both able to exploit an increase in the number of receive
antennas, as well as in the case of low SNR. Finally IGMM rule, exploiting
a second-order characterization of the received vector under both
hypotheses, has the most appealing performance when considering all
the three scenarios.
\end{itemize}
Successively, we considered a scenario with a (possibly distributed)
``Rician'' jamming interference and tackled the resulting composite
hypothesis testing problem within the GLRT framework. More specifically,
we developed sub-optimal GLRT-like decision rules which extend IS,
NLOS and IGMM rules to the case of subspace interference. With reference
to these rules, the following trends have been observed:
\begin{itemize}
\item All the considered ``interference-aware'' rules (IS-GLRT, NLOS-GLRT
and IGMM-GLRT) significantly outperform the ``interference-unaware''
counterparts in the case of a moderate-to-high SNR level and non-negligible
LOS condition, as in such case system performance is interference-dominated
and thus interference suppression leads to a remarkable gain. Also,
it has been shown that all these rules benefit from increase of $N$
for enhanced interference suppression, with the sole exception of
IGMM-GLRT in a NLOS case (due to lack of identifiability). Numerical
evidence has also underlined the appeal of IGMM-GLRT and IS-GLRT in
LOS and Intermediate/NLOS setups, respectively;
\end{itemize}
Finally, asymptotic equivalences established among all these rules
in the case of either interference-free or interference-prone scenarios
were confirmed by simulations. Future research tracks will concern
theoretical performance analysis of the proposed rules and design
of advanced fusion schemes robust to smarter jammers.

\appendix{}

In this appendix we will provide the second-order characterization
of $\bm{y}|\mathcal{H}_{i}$\label{sec: Appendix Second order characterization-1}.
The mean vector $\mathbb{E}\{\bm{y}|\mathcal{H}_{i}\}$ is evaluated
as:
\begin{gather}
\mathbb{E}\{\bm{y}|\mathcal{H}_{i}\}=\mathbb{E}\{\bar{\bm{H}}\bm{D}^{1/2}\bm{x}+\bm{w}|\mathcal{H}_{i}\}=\\
\mathbb{E}\{\bar{\bm{H}}\}\,\bm{D}^{1/2}\,\mathbb{E}\{\bm{x}|\mathcal{H}_{i}\}=\widetilde{\bm{A}}(\bm{\theta})\,\bm{\rho}_{i}\label{eq:steering_mat_def}
\end{gather}
where we have exploited $\mathbb{E}\{\bm{w}\}=\bm{0}_{N}$ and statistical
independence between fading coefficients and sensors decisions, respectively.
Finally in Eq. (\ref{eq:steering_mat_def}) we have recalled the definitions
of matrix $\widetilde{\bm{A}}(\bm{\theta})$, whose $k$th column
equals $\bm{\mu}_{k}=b_{k}\,\sqrt{\beta_{k}}\,\bm{a}(\theta_{k})$,
and of $\bm{\rho}_{i}=\begin{bmatrix}P_{i,1} & \cdots & P_{i,K}\end{bmatrix}^{T}$.

Differently, the covariance matrix is expressed as:
\begin{gather}
\bm{\Sigma}_{\bm{y}|\mathcal{H}_{i}}=\nonumber \\
\mathbb{E}\{(\bm{y}-\widetilde{\bm{A}}(\bm{\theta})\,\bm{\rho}_{i})\,(\bm{y}-\widetilde{\bm{A}}(\bm{\theta})\,\bm{\rho}_{i})^{\dagger}\,|\mathcal{H}_{i}\}=\\
\widetilde{\bm{A}}(\bm{\theta})\,\mathbb{E}\{(\bm{x}-\bm{\rho}_{i})(\bm{x}-\bm{\rho}_{i})^{T}\,|\mathcal{H}_{i}\}\,\widetilde{\bm{A}}(\bm{\theta})^{\dagger}\,+\nonumber \\
\mathbb{E}\{(\bm{H}\,\bm{B}_{s}\,\bm{x})(\bm{H}\,\bm{B}_{\mathrm{s}}\,\bm{x})^{\dagger}\,|\mathcal{H}_{i}\}+\mathbb{E}\{\bm{w}\bm{w}^{\dagger}\}=\\
\widetilde{\bm{A}}(\bm{\theta})\,\bm{\Sigma}_{\bm{x}|\mathcal{H}_{i}}\,\widetilde{\bm{A}}(\bm{\theta})^{\dagger}+\nonumber \\
\mathbb{E}\left\{ \left(\sum_{k=1}^{K}\bm{h}_{k}\,\sqrt{\nu_{k}}\,x_{k}\right)\left(\sum_{\ell=1}^{K}\bm{h}_{\ell}^{\dagger}\,\sqrt{\nu_{\ell}}\,x_{\ell}\right)|\mathcal{H}_{i}\right\} +\sigma_{w}^{2}\,\bm{I}_{N}\label{eq: Cov. pre-final}
\end{gather}
where $\bm{B}_{s}\triangleq(\bm{I}_{K}-\bm{R}^{2})^{1/2}\bm{D}^{1/2}$
and we recall $\nu_{k}=(1-b_{k}^{2})\beta_{k}$. The second term in
Eq. (\ref{eq: Cov. pre-final}) can be simplified as
\begin{gather}
\mathbb{E}\left\{ \left(\sum_{k=1}^{K}\bm{h}_{k}\,\sqrt{\nu_{k}}\,x_{k}\right)\left(\sum_{\ell=1}^{K}\bm{h}_{\ell}^{\dagger}\,\sqrt{\nu_{\ell}}\,x_{\ell}\right)|\mathcal{H}_{i}\right\} =\nonumber \\
\sum_{k=1}^{K}\mathbb{E}\{\bm{h}_{k}\,\bm{h}_{k}^{\dagger}\}\,\nu_{k}\,\mathbb{E}\{x_{k}^{2}|\mathcal{H}_{i}\}=\sum_{k=1}^{K}\nu_{k}\,P_{i,k}\,\bm{I}_{N}\label{eq: cov_term_Exp}
\end{gather}
which follows from mutual independence of vectors $\bm{h}_{k}$, $k\in\mathcal{K}$.
Then, substituting back Eq. (\ref{eq: cov_term_Exp}) in Eq. (\ref{eq: Cov. pre-final})
gives:
\begin{gather}
\bm{\Sigma}_{\bm{y}|\mathcal{H}_{i}}=\widetilde{\bm{A}}(\bm{\theta})\,\bm{\Sigma}_{\bm{x}|\mathcal{H}_{i}}\,\widetilde{\bm{A}}(\bm{\theta})^{\dagger}+\sigma_{e,i}^{2}\,\bm{I}_{N}
\end{gather}
where $\sigma_{e,i}^{2}\triangleq[\sum_{k=1}^{K}\nu_{k}\,P_{i,k}+\sigma_{w}^{2}]$.
Analogously, we can evaluate the pseudo-covariance of $\bm{y}|\mathcal{H}_{i}$
as
\begin{gather}
\bar{\bm{\Sigma}}_{\bm{y}|\mathcal{H}_{i}}=\mathbb{E}\{(\bm{y}-\widetilde{\bm{A}}(\bm{\theta})\,\bm{\rho}_{i})\,(\bm{y}-\widetilde{\bm{A}}(\bm{\theta})\,\bm{\rho}_{i})^{T}\,|\mathcal{H}_{i}\}\\
=\widetilde{\bm{A}}(\bm{\theta})\,\mathbb{E}\{(\bm{x}-\bm{\rho}_{i})(\bm{x}-\bm{\rho}_{i})^{T}\,|\mathcal{H}_{i}\}\,\widetilde{\bm{A}}(\bm{\theta})^{T}\,+\nonumber \\
\mathbb{E}\{(\bm{H}\,\bm{B}_{\mathrm{s}}\,\bm{x})(\bm{H}\,\bm{B}_{\mathrm{s}}\,\bm{x})^{T}\,|\mathcal{H}_{i}\}\\
=\widetilde{\bm{A}}(\bm{\theta})\,\bm{\Sigma}_{\bm{x}|\mathcal{H}_{i}}\,\widetilde{\bm{A}}(\bm{\theta})^{T}+\nonumber \\
\mathbb{E}\left\{ \left(\sum_{k=1}^{K}\bm{h}_{k}\,\sqrt{\nu_{k}}\,x_{k}\right)\left(\sum_{\ell=1}^{K}\bm{h}_{\ell}^{T}\,\sqrt{\nu_{\ell}}\,x_{\ell}\right)|\mathcal{H}_{i}\right\} \label{eq: pseudocov_term_exp}
\end{gather}
since $\mathbb{E}\{\bm{w}\,\bm{w}^{T}\}=\bm{O}_{N}$ (i.e., the noise
is assumed circular). Also, it can be shown that the second term in
Eq. (\ref{eq: pseudocov_term_exp}) is a null matrix, since
\begin{gather}
\mathbb{E}\left\{ \left(\sum_{k=1}^{K}\bm{h}_{k}\,\sqrt{\nu_{k}}\,x_{k}\right)\left(\sum_{\ell=1}^{K}\bm{h}_{\ell}^{T}\,\sqrt{\nu_{\ell}}\,x_{\ell}\right)|\mathcal{H}_{i}\right\} =\nonumber \\
\sum_{k=1}^{K}\mathbb{E}\{\bm{h}_{k}\,\bm{h}_{k}^{T}\}\,\nu_{k}\,\mathbb{E}\{x_{k}^{2}|\mathcal{H}_{i}\}=\bm{O}_{N}
\end{gather}
since the NLOS fading vector $\bm{h}_{k}$ is assumed circular. Therefore
the final expression for the pseudo-covariance is:
\begin{equation}
\bar{\bm{\Sigma}}_{\bm{y}|\mathcal{H}_{i}}=\widetilde{\bm{A}}(\bm{\theta})\,\bm{\Sigma}_{\bm{x}|\mathcal{H}_{i}}\,\widetilde{\bm{A}}(\bm{\theta})^{T}\label{eq: pseudocovariance final}
\end{equation}
Eq. (\ref{eq: pseudocovariance final}) is not a null matrix, thus
motivating augmented form processing.

\bibliographystyle{IEEEtran}
\bibliography{IEEEabrv,sensor_networks}

\end{document}